%% file: main_nips24_fgbert.tex
\documentclass{article}

\usepackage[preprint,nonatbib]{neurips_2024}

\usepackage[utf8]{inputenc} 
\usepackage[T1]{fontenc}    
\usepackage{hyperref}       
\usepackage{url}            
\usepackage{booktabs}       
\usepackage{amsfonts}       
\usepackage{nicefrac}       
\usepackage{microtype}      
\usepackage{xcolor}         

\definecolor{mycolor1}{RGB}{31, 119, 180}
\definecolor{mycolor2}{RGB}{255, 127, 14}

\usepackage{longtable}
\usepackage{ltablex}
\usepackage{array} 
\usepackage{booktabs} 
\usepackage{tabularx} 
\usepackage{threeparttable}
\usepackage{graphicx} 
\usepackage{subcaption} 
\usepackage{xcolor}
\usepackage{color}
\usepackage{caption}
\usepackage{subcaption}
\usepackage{makecell}
\usepackage{multirow}
\usepackage{amsmath}
\usepackage{verbatim}
\usepackage{colortbl}
\usepackage{pifont}
\usepackage{adjustbox}
\usepackage[normalem]{ulem}
\usepackage{wrapfig}

\title{{FGBERT}: Function-Driven Pre-trained Gene Language Model for Metagenomics}

\author{
    Chenrui Duan$^{1,2}$~
    Zelin Zang$^{2}$~
    Yongjie Xu$^{1,2}$~
    Hang He$^{2}$~
    Zihan Liu$^{1,2}$~
    Siyuan Li$^{1,2}$~\\
    \textbf{Zijia Song}$^{3}$~
    \textbf{Jusheng Zheng}$^{2}$~
    \textbf{Stan Z. Li}$^{2\dag}$\\
    $^{1}$AI Lab, Research Center for Industries of the Future, Westlake University;\\
    $^{2}$Zhejiang University; \quad
    $^{3}$National University of Defense Technology \\
    \texttt{duanchenrui@westlake.edu.cn}; \\
    \{\texttt{zangzelin};~\texttt{xuyongjie};~\texttt{hehang};~\texttt{liuzihan};~\texttt{lisiyuan};\}\texttt{@westlake.edu.cn}\\
    \texttt{songzijia}\texttt{@nudt.edu.cn}\\
    \{~\texttt{zhengjusheng};~\texttt{stan.zq.li};\}\texttt{@westlake.edu.cn}\\
}
 
\begin{document}



\maketitle
\thispagestyle{empty}

\maketitle

\begin{abstract}
Metagenomic data, comprising mixed multi-species genomes, are prevalent in diverse environments like oceans and soils, significantly impacting human health and ecological functions. 
However, current research relies on K-mer, which limits the capture of structurally and functionally relevant gene contexts. Moreover, these approaches struggle with encoding biologically meaningful genes and fail to address the One-to-Many and Many-to-One relationships inherent in metagenomic data. 
To overcome these challenges, we introduce {FGBERT}, a novel metagenomic pre-trained model that employs a protein-based gene representation as a context-aware and structure-relevant tokenizer. {FGBERT} incorporates Masked Gene Modeling (MGM) to enhance the understanding of inter-gene contextual relationships and Triplet Enhanced Metagenomic Contrastive Learning (TMC) to elucidate gene sequence-function relationships. 
Pre-trained on over 100 million metagenomic sequences, {FGBERT} demonstrates superior performance on metagenomic datasets at four levels, spanning gene, functional, bacterial, and environmental levels and ranging from 1k to 213k input sequences. 
Case studies of ATP Synthase and Gene Operons highlight {FGBERT}'s capability for functional recognition and its biological relevance in metagenomic research.
\end{abstract}


\section{Introduction}\label{introduction}
\input{sec_intro}

\section{Related Works}\label{relatedworks}
\input{sec_relatedwork}

\section{Methods}\label{methods}
\input{sec_methods}

\section{Experiments}\label{experiments}
\input{sec_experiments}

\section{Conclusion}\label{conclusion}
\input{sec_conclusion}

\bibliographystyle{plain}
\bibliography{example_paper}

\section{Appendix}\label{appendix}
\input{sec_appendix}

\end{document}

%% file: sec_intro.tex
Metagenomics, the study of mixed genomes of microbial communities in the environment (e.g. gut microbiomes or soil ecosystems)~\cite{mande2012classification,mathieu2022machine}, has revealed the critical role in fundamental biological processes like enzyme synthesis, gene expression regulation, and immune function~\cite{pavlopoulos2023unraveling}. 
This deepened understanding highlights the need to accurately interpret the intricate genetic information contained within these diverse communities.
Consequently, deciphering the complex sequences of multiple species in metagenomics is vital for unravelling life's mechanisms and advancing biotechnology~\cite{albertsen2023long,lin2023evolutionary,liu2022opportunities}. 

\begin{figure}[t]
    \vspace{-2em}
    \centering
    \includegraphics[width=1.0\textwidth]{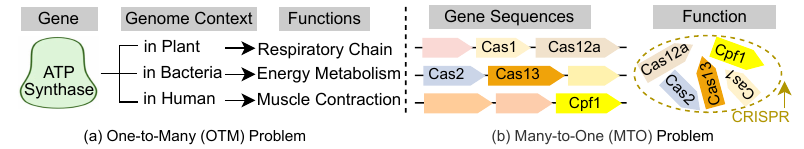}
    \vspace{-0.5em}
    \caption{Motivaion. Two types of complex relationships between gene sequences and functions in metagenomics. \textbf{One-to-Many} problem means that the same gene may display different functions based on the genomic context; for example, ATP synthase works differently in plants, heterotrophic bacteria, and humans. \textbf{Many-to-One} problem shows that multiple genes may perform the same function; for instance, different genes from different bacteria, e.g., Cpf1, Cas1, etc., produce the same resistance function within the immune system CRISPR.}
    \label{fig:figure1}
    \vspace{-1.5em}
\end{figure}

Unlike traditional genomics focused on single species, metagenomics involves genetic material directly from environmental samples, posing significant challenges due to sample diversity and species abundance~\cite{lu2022machine}. As shown in Fig.~\ref{fig:figure1}, the typical challenges in metagenomics are the presence of One-to-Many (OTM) and Many-to-One (MTO) problems.
The OTM problem indicates that a single gene can exhibit various functions in different genomic contexts, underscoring the significance of inter-gene interactions in function regulation~\cite{yang2021review}. 
For example, ATP synthase displays distinct functionalities in diverse organisms such as bacteria, plants, and humans (Fig.~\ref{fig:figure1}a).
Conversely, the MTO problem implies that different genes can share the same function, emphasizing expression commonality~\cite{al2022diverse}.  
For example, the CRISPR immune mechanism involves various proteins like Cpf1, Cas1, and Cas13, each contributing to the same defensive function (Fig.~\ref{fig:figure1}b)~\cite{yang2021review, al2022diverse, hu2022metagenomic}.

Recently, various computational methods have emerged for genomic and metagenomic data analysis. However, these methods still face challenges when analyzing metagenomic data.
Firstly, the \textbf{Semantic Tokenizer} problem.
Most machine learning-based methods~\cite{hoarfrost2022deep,liang2020deepmicrobes,miller2022deciphering}, taking K-mer counts, frequencies or embeddings as input features, providing efficient alternatives to traditional homology searches against reference genome databases. However, K-mer features often have limited representation ability and fail to capture global information. 
Secondly, the \textbf{Function-Driven Modeling} problem.
Although recent Transformer models excel in modeling complex DNA contexts through long-range dependencies, they are predominantly tailored for single-species genomic analysis~\cite{wolf2020transformers, zhou2023dnabert, dalla2023nucleotide} and do not adequately address OTM and MTO challenges. This limitation impedes their ability to accurately model the intricate relationships between genes, their function across diverse genomic environments, and their connections among sequences with similar functions.
Thirdly, the \textbf{Low Generalization} problem. Models like MetaTransformer~\cite{wichmann2023metatransformer} and ViBE~\cite{gwak2022vibe}, designed for specific tasks such as read classification and virus category prediction, fail to grasp the broader biological complexities of metagenomic data, limiting their generalization across diverse metagenomic tasks.

To address these challenges, we propose {FGBERT}, a novel metagenomic pre-trained model designed to encode contextually-aware and functionally relevant representations of metagenomic sequences.
First, to solve the problem of encoding gene sequences with biological meaning, we propose a protein-based gene representation as a context-aware tokenizer, allowing for a flexible token vocabulary for longer metagenomic sequences. This strategy leverages the inherent protein functional and structural information encoded in metagenomic data~\cite{pavlopoulos2023unraveling}, overcoming the limitations of K-mer methods and maintaining functional consistency despite potential mutations~\cite{d2014redundancy}. 
Second, we propose two pre-training tasks for function-driven modeling: Masked Gene Modeling (MGM) and Triplet Enhanced Metagenomic Contrastive Learning (TMC) to enhance the co-representation learning of metagenomic gene sequences and functions.
Thirdly, {FGBERT} is pre-trained on over 100 million sequences, showcasing robust performance across diverse datasets spanning gene, functional, bacterial, and environmental levels. 

\textbf{Contributions.} In this work, we identify three key challenges in metagenomic analysis. To address these issues, we propose {FGBERT}. To the best of our knowledge, this is the first metagenomic pre-trained model encoding context-aware and function-relevant representations of metagenomic sequences. To summarize: (1) We introduce a new idea of protein-based gene representations to learn biologically meaningful tokenization of long sequences. (2) We propose MGM to model inter-gene relationships and TMC to learn complex relationships between gene sequences and functions. (3) We conduct extensive experiments across various downstream tasks, spanning gene, functional, bacterial, and environmental levels with input sizes from 1k to 213k sequences. {FGBERT} achieves SOTA performance.

\begin{figure*}
    \centering        
    \vspace{-2em}
    \includegraphics[width=\textwidth]{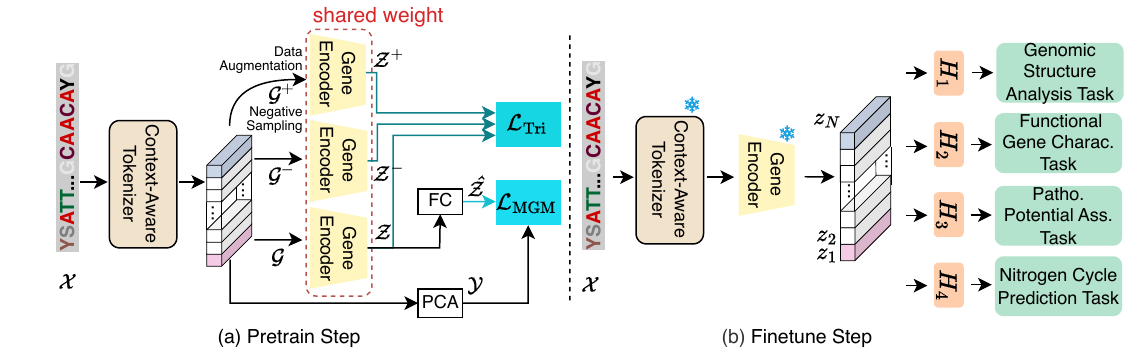}
    \vspace{-0.5em}
    \caption{Overview of {FGBERT}.
    A metagenomic sequence $\mathcal{X}$ is converted into ordered protein-based gene representations $\mathcal{G}$ via a Context-Aware Tokenizer.
    Next, we pre-train a Gene Encoder with $\mathcal{L}_{\text{MGM}}$, 15\% of these tokens are masked to predict labels $\mathcal{Y}$.
    Meanwhile, we introduce $\mathcal{L}_{\text{Tri}}$ to distinguish gene sequences. The data augmentation and negative sampling modules generate positive samples $\mathcal{G}^+$ and negative samples $\mathcal{G}^-$, respectively. 
    Finally, after fine-tuning, {FGBERT} can handle various downstream tasks.
    }    
    \label{fig:figure2}
    \vspace{-0.5cm}    
\end{figure*}

%% file: sec_relatedwork.tex
\textbf{Research on Metagenomics.}
Traditional alignment-based methods like MetaPhlAn5~\cite{segata2012metagenomic} aim to match similarities between query sequences and known reference genomes and are common for taxonomic profiling. 
Advancements in deep learning have led to new methods like CNN-MGP~\cite{al2019cnn} and DeepVirFinder~\cite{ren2020identifying}, which use CNNs for gene and viral classifications with one-hot encoding. 
K-mer tokenization~\cite{fiannaca2018deep}, employed in approaches like MDL4Microbiome~\cite{lee2022multimodal}, is a standard for DNA sequence characterization.
Additionally, Virtifier~\cite{miao2022virtifier} maps a nucleotide sequence using a codon dictionary combined with LSTM to predict viral genes.
DeepMicrobes~\cite{liang2020deepmicrobes} employs a self-attention mechanism, while DeepTE~\cite{yan2020deepte} uses K-mer inputs with CNNs for element classification, and Genomic-nlp~\cite{miller2022deciphering} applies word2vec for gene function analysis.
MetaTransformer~\cite{wichmann2023metatransformer} uses K-mer embedding for species classification with Transformer. For pre-training models, LookingGlass~\cite{hoarfrost2022deep} uses a three-layer LSTM model for functional prediction in short DNA reads. ViBE~\cite{gwak2022vibe} employs a K-mer token-based BERT model pre-trained with Masked Language Modeling for virus identification.

\textbf{Pre-Training on Genomics.}
The BERT model, effective in DNA sequence characterization, is limited by the Transformer architecture's computational burden. 
LOGO~\cite{yang2022integrating} addresses this by cutting off long sequences into 1-2kb sub-sequences.
Enformer~\cite{avsec2021effective} combines extended convolution with Transformers for long human genomic data. GenSLMs~\cite{zvyagin2022genslms} introduce hierarchical language models for whole-genome modelling. 
DNABERT~\cite{ji2021dnabert}, the first pre-trained model on the human genome that focuses on extracting efficient genomic representations. DNABERT2~\cite{zhou2023dnabert}, its successor, uses Byte Pair Encoding on multi-species genomic data.
NT~\cite{dalla2023nucleotide} is trained on nucleotide sequences from humans and other species and evaluated on 18 genome prediction tasks. HyenaDNA~\cite{nguyen2023hyenadna} presents a long-range genomic model based on single-nucleotide polymorphisms on human reference genomes.

%% file: sec_methods.tex
In this section, we provide a detailed description of the proposed pre-training model {FGBERT}, which contains the MGM and TMC components as depicted in Fig.~\ref{fig:figure2}.

\textbf{Notation.}
Given a dataset of metagenomic long sequences $\{\mathcal{X}_i\} _{i=1}^m$, we simplify each $\mathcal{X}_i$ into a set of shorter gene sequences $\{{x_i}\}_{i=1}^{n_{i}}$ using ultrasonic fragmentation and assembly techniques~\cite{KUSTERS19934119}, where $n_{i}$ represents the variable number of gene sequences derived from each $\mathcal{X}_i$. Each gene sequence $x_i$ is tokenized in a vector $g_i \in \mathbb{R}^{d}$, where $d$ is the token dimension. Each sequence is associated with a reduced-dimensional representation $y_i\in \mathbb{R}^{100}$. Suppose a gene group $\mathcal{G}={\left \{ g_i \right \} }_{i=1}^N$ is formed by concatenating $N$ gene vectors sequentially. During the pre-training phase, each gene token $g_i\in \mathcal{G}$ is processed by the context-aware genome language encoder $\mathcal{F}(\cdot)$, generating the knowledge representations $z_i$, where $z_i=\mathcal{F}(g_i)$. These representations are then transformed by a fully connected layer into $\hat{z}_i$, defined as $\hat{z}_i=\mathcal{H}(z_i)$, where $\mathcal{H(\cdot)}$ represents a multi-classification head.
In addition, we incorporate contrastive learning into the methodology. For each gene $x_i$, we introduce a data augmentation module to generate positive samples $x_{j(i)}$ and a hard negative sampling strategy for constructing negative samples $x_{k(i)}$.

\subsection{Context-aware Masked Gene Modeling~(MGM) for One-to-Many problem}

\textbf{Context-Aware Tokenizer.}
To develop a biologically meaningful tokenization of long sequences, we design a context-aware tokenizer utilizing the Protein Language Model (PLM), such as ESM-2~\cite{lin2022language} with 15 billion parameters, integrating biological prior knowledge.
As illustrated in Fig.~\ref{fig:figure3}, this tokenizer framework begins by extracting DNA gene sequences $\{{x_i}\}_{i=1}^{n_{i}}$ from a metagenomic sequence $\mathcal{X}_i$ utilizing the European Nucleotide Archive (ENA) software~\cite{gruenstaeudl2020annonex2embl}. 
This conversion enhances the flexibility of analyzing longer metagenomic sequences. 

\begin{wraptable}{r}{0.6\textwidth}
    \vspace{-1.2em}
    \centering
    \includegraphics[width=\linewidth]{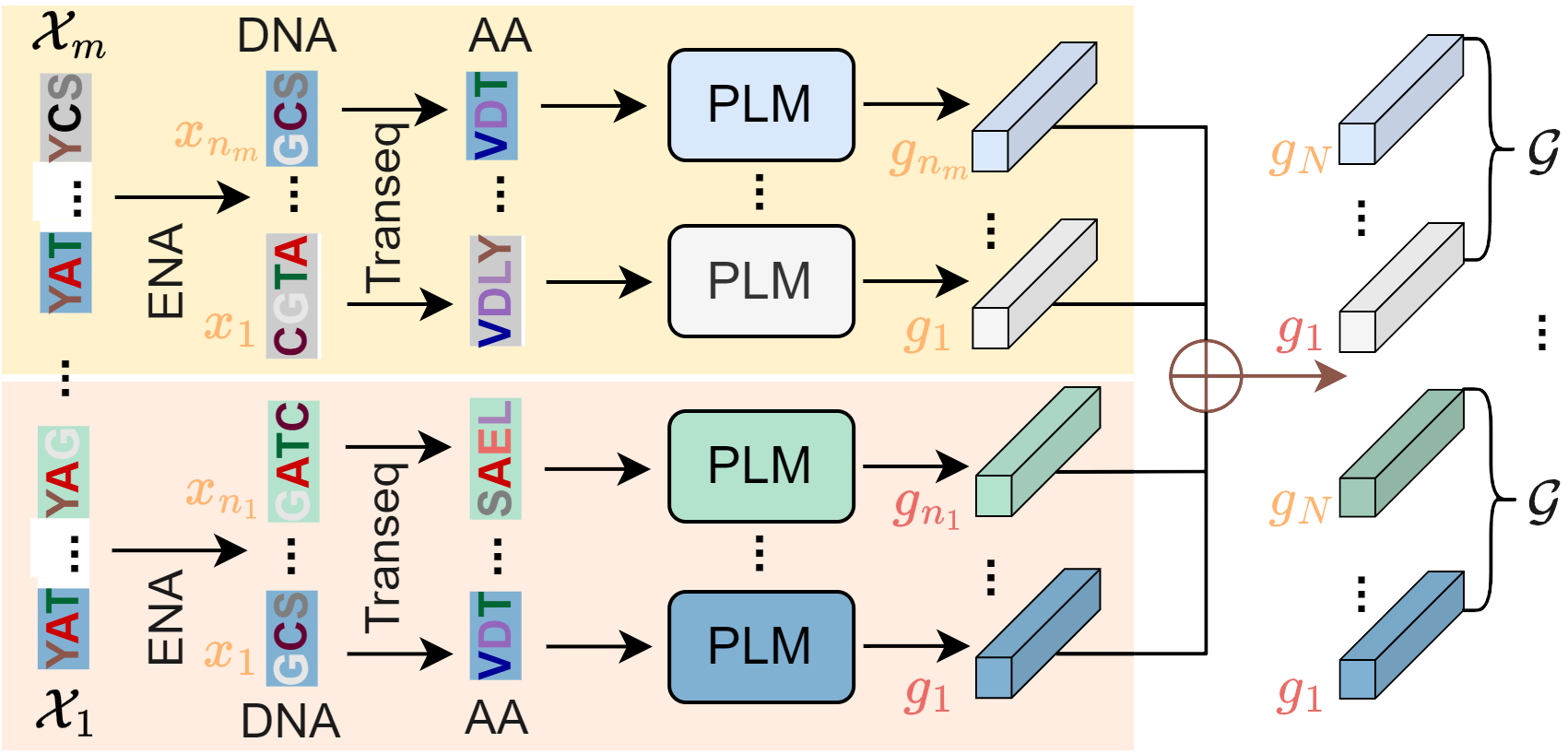}   
    \vspace{-0.75em}
    \caption{Framework of Context-Aware Tokenizer.
        }
    \label{fig:figure3}
    \vspace{-1.2em}
\end{wraptable}

Secondly, these DNA sequences $x_i$ are translated into Amino Acid (AA) sequences using the Transeq software~\cite{mcwilliam2013analysis}. This translation addresses the issue of degenerate DNA codes, where certain non-`ATCG' symbols like `Y' or `S' can represent multiple nucleotides (e.g., `Y' can be `C' or `G'). By translating to AA sequences, we eliminate this redundancy, as different DNA sequences can map to the same AA sequence, ensuring consistency in representing biologically equivalent sequences~\cite{lawson2004catabolite,jain2017duplication}.
Thirdly, these AA sequences are transformed into 1280D normalized ESM-2 representations, with an additional gene orientation vector, resulting in 1281D gene representations $\{g_i\}_{i=1}^{n_{i}}$. 
The utility of ESM-2 lies in its ability to reveal gene relationships and functional information inherent in metagenomic data~\cite{pavlopoulos2023unraveling}, preserving important intra-gene contextually-aware information.
Finally, these representations every $N$ representations are concatenated sequentially to form gene groups $\mathcal{G}$, which serve as the basis for subsequent modeling tasks. 


\textbf{Masked Gene Modeling.}
We propose the MGM to enhance the model’s understanding of the relationships between genes within metagenomic sequences and their function regulations across diverse genomic environments (OTM problem).
During pre-training, each gene token is masked with a 15\% probability and predicted based on its unmasked genome context ${\mathcal{G}_{/M}}$:
\vspace{-0.5em}
\begin{equation}
    \mathcal{L}_{\text{MLM}} = \mathbb{E}_{g\sim \mathcal{G}} \mathbb{E}_M\sum_{i \in M}- \log p(g_i|{\mathcal{G}_{/M}}),
\end{equation}
where $M$ denotes the index set of the masked gene.

In addition, considering genetic polymorphism~\cite{pastinen2006influence,zhang2021metagenomics}, the MGM component focuses on detecting multiple genes that could coexist at a single genomic site, denoted as $\hat{z}_i=[\hat{z}_1,\hat{z}_2,\hat{z}_3,\hat{z}_4]$. This requires the model not only to predict individual genes but also to identify various combinations of genes occurring within the same site.
Thus, we enhance the model with a comprehensive loss function, $\mathcal{L}_{\text{MGM}}$, which incorporates feature reconstruction and probability prediction:
\vspace{-0.5em}
\begin{equation}
    \mathcal{L}_{\text{MGM}} = \frac{1}{N}\textstyle\sum\limits_{i=1}^{N}{( 1-\frac{y_i^T \hat{z}_i}{\left \| y_i \right\|\cdot\left\| \hat{z}_i\right\|})}^{\gamma}+ \\ \frac{\alpha}{N}\textstyle\sum\limits_{i=1}^{N} || \tilde{z}_i - \tilde{y}_i||_2^2,
\end{equation}
where $\gamma$ is a reconstruction loss with the scaled cosine error, and $\alpha$ is a weighting factor to balance the importance of the two loss functions. 
The first item, Feature Reconstruction Loss (FRL), quantifies the distance between the model prediction $\hat{z}_i$ and its corresponding label $y_i$. The second item, Probability Prediction Loss (PPL), evaluates the discrepancy between the predicted embedding probability $\tilde{z}_i=\frac{e^{\hat{z}_i}}{\sum_{j=1}^{C}e^{\hat{z}_j}} $ and the true category probability $\tilde{y}_i = \frac{e^{y_i}}{\sum_{j=1}^{C}e^{y_j}}$, both processed via the softmax function. $C$ denotes the number of gene combinations and is set to 4.

\subsection{Triplet Metagenomic Contrastive Framework~(TMC) for Many-to-One problem}
\textbf{Contrastive Learning.}
The MGM component enhances the model's ability to learn contextual relationships between genes, helping to alleviate the OTM problem present in metagenomic data. However, when faced with the MTO problem, the model's ability to describe the relationships between sequences with the same function remains weak. For instance, when annotating Enzyme Commission (EC) numbers for partial metagenomic sequences, we observe that sequences with the same EC number are close to each other in feature space, while sequences with different EC numbers are farther apart~\cite{jumper2021highly}. Therefore, we introduce a contrastive learning technique to capture the functional relationships between gene classes, enabling different genes with similar functions to cluster together and further optimize model training. Generally speaking, the objective of contrastive learning is to learn an embedding function $\mathcal{F}$ such that the distance between positive pairs is smaller than the distance between negative pairs: 
\vspace{-0.75em}
\begin{equation}
d(\mathcal{F}(x_a),\mathcal{F}(x_p)) < d(\mathcal{F}(x_a),\mathcal{F}(x_n)),
\end{equation}
where $d(\cdot,\cdot)$ is a distance function (e.g., Euclidean distance) defined on the embedding space. We adopt SupCon-Hard loss~\cite{khosla2020supervised} to consider multiple positive and negative samples for each anchor, encouraging the model to mine difficult samples and enhancing its robustness.
Additionally, data augmentation and negative sampling modules are included to create positive and negative samples, further improving the model's capacity to recognize commonalities among gene classes.

\textbf{Positives Sampling.} The strategy for sampling triplets is crucial to learn a well-organized embedding space.
For each gene group $\mathcal{G}$, as an anchor gene $x_i$ within a gene batch $I$, a mutation strategy is proposed to augment orphan sequences (i.e., functions associated with individual sequences) to generate a large number of pairs of positive samples $x_{j(i)}\in \mathcal{G}_i^+$, where $\mathcal{G}_i^+$ is the set of positive samples for anchor $x_i$. 
Specifically, 10 random mutations are performed for each gene sequence, with mutation ratios randomly generated according to a standard normal distribution. The number of mutations is calculated based on sequence lengths. 
This process aims to generate new sequences that are functionally similar to the original sequence but sequentially different, providing additional training data to improve the predictive power and accuracy of orphan EC numbers.

\textbf{Hard Negatives Sampling}
Previous studies~\cite{hermans2017defense} have demonstrated that a critical component of successful contrastive learning is balancing the triviality and hardness of the sampled triplets. For the negative sample pair $x_{k(i)}\in \mathcal{G}_i^-$, where $\mathcal{G}_i^-$ represents the set of negative samples for the anchor $x_i$, this balance is particularly important.
We determine the centers for each functional group by averaging the embeddings of all sequences within that group. Subsequently, we compute the Euclidean distances $d(\cdot)$ based on these centers. For negative sample selection, we choose samples that are similar to the anchor in latent space but from different clusters, thus increasing the learning difficulty compared to random selection.
The triplet loss $ \mathcal{L}_{\text{Tri}}( x_i, \{ x_{j(i)} \}_{j=1}^{N_j}, \{ x_{k(i)} \}_{k=1}^{N_k})$ is defined:
\vspace{-0.5em}
\begin{equation}
    \mathcal{L}_{\text{Tri}} = -\sum_{i\in I}\log( {1}/{|\mathcal{G}_i^+|} \sum_{j \in \mathcal{G}_i^+} {\exp(S_{z_i, z_{j(i)}}/\tau)}/{\mathcal{G}_i}).
\end{equation}
For all negative samples in $\mathcal{G}_i^-$, the probability of selecting each negative sample $x_k$ for the anchor $x_i$ as follows:
\vspace{-0.25em}
\begin{equation}
    \mathcal{G}_i = \sum_{x_{j(i)}} \exp(S_{z_i, z_{j(i)}} / \tau) + \sum_{x_{k(i)}} p_{x_k} \exp(S_{z_i, z_{k(i)}} / \tau),
\end{equation}
where $\tau$ is the temperature hyper-parameter, and $S$ is the similarity function, typically cosine similarity. The term $p_{x_k}$ represents the probability of selecting the negative sample $x_k$ for the anchor $x_i$, calculated as $p_{x_k} = {w_{x_k}}/{\sum_{x_m \in \mathcal{G}_i^-}w_{x_m}}$ with $w_{x_k} = \frac{1}{d(x_i, x_k)} $.

Finally, MGM and TMC constitute a unified pre-training framework with a total loss:
\vspace{-0.25em}
\begin{equation}
    \mathcal{L}_{\text{Total}} = \mathcal{L}_{\text{MGM}} + \lambda \mathcal{L}_{\text{Tri}}, 
\end{equation}
where $\lambda$ is a hyper-parameter tuning the influence between two loss functions.

%% file: sec_experiments.tex
\vspace{-1em}

\begin{wrapfigure}{r}{0.62\textwidth}
    \vspace{-2.5em}
    \centering
    \captionof{table}{Description of Experimental Datasets.}
    \label{tab:experiment datasets}
    \footnotesize
    \vspace{-0.5em}
    \begin{adjustbox}{width=\linewidth}
    \begin{tabular}{lllll} 
    \toprule
    \textbf{Task}                                                                                               & \textbf{Dataset} & \textbf{Description} & \textbf{\#Seq.} & \textbf{\#Class}  \\ 
    \midrule
    \begin{tabular}[c]{@{}l@{}}Gene Structure Prediction\\\textbf{(Gene Level)}\end{tabular}                    & E-K12            & Gene Operons         & 4,315             & 1,379             \\ 
    \midrule
    \multirow{5}{*}{\begin{tabular}[c]{@{}l@{}}Functional Prediction\\\textbf{(Functional Level)}\end{tabular}} & CARD-A           & AMR Family           & 1,966             & 269               \\
                                                                                                                & CARD-D           & Drug Class           & 1,966             & 37                \\
                                                                                                                & CARD-R           & Resistance Mech.     & 1,966             & 7                 \\
                                                                                                                & VFDB             & Virulence Fact.      & 8,945             & 15                \\
                                                                                                                & ENZYME           & Enzyme Func.         & 5,761             & 7                 \\ 
    \midrule
    \begin{tabular}[c]{@{}l@{}}Pathogenicity Prediction \\\textbf{(Bacteria Level) }\end{tabular}               & PATRIC           & Pathogenic Genes     & 5,000             & 110               \\ 
    \midrule
    \begin{tabular}[c]{@{}l@{}}Nitrogen Cycle Prediction \\\textbf{(Environmental Level) }\end{tabular}         & NCycDB           & Cycling Genes        & 213,501           & 68                \\
    \bottomrule
    \end{tabular}
    \end{adjustbox}
    \vspace{-1.3em}
\end{wrapfigure}

We pre-train {FGBERT} on a large amount of metagenomic data and comprehensively assess its generalization on different datasets ranging from thousands to hundreds of thousands of sequences, as detailed in Tab.~\ref{tab:experiment datasets}.
Our model is tested across four task levels: (1) Gene Structure Analysis, (2) Functional Gene Prediction, (3) Pathogenicity Potential Assessment, and (4) Nitrogen Cycle Prediction.
More detailed description of the downstream tasks can be found in Appendix Sec.~\ref{app:overview of downstream}.
We use the MGnify database (updated February 2023), which comprises 2,973,257,435 protein sequences from various microbial communities, detailed in Appendix Tab.~\ref{tab:MGnify_catalogues}.

\subsection{Experiments Results on Four Level Downstream Tasks}
\textbf{Level 1 Task A: Gene Structure Analysis $\rightarrow$   Gene Operons Prediction.}
This task is to identify the transcription factor binding sites that are strongly correlated with operon regulation in the gene regulatory network, which helps us to understand the mechanism and network of gene regulation.
The dataset used is the E. coli K12 RegulonDB dataset (E-K12)~\cite{salgado2018using}, which contains 4315 operons.
Detailed information is listed in Appendix Tab.~\ref{tab:operons_dataset}.

\begin{wraptable}{r}{0.43\textwidth}
    \vspace{-2.5em}
    \centering
    \includegraphics[width=\linewidth]{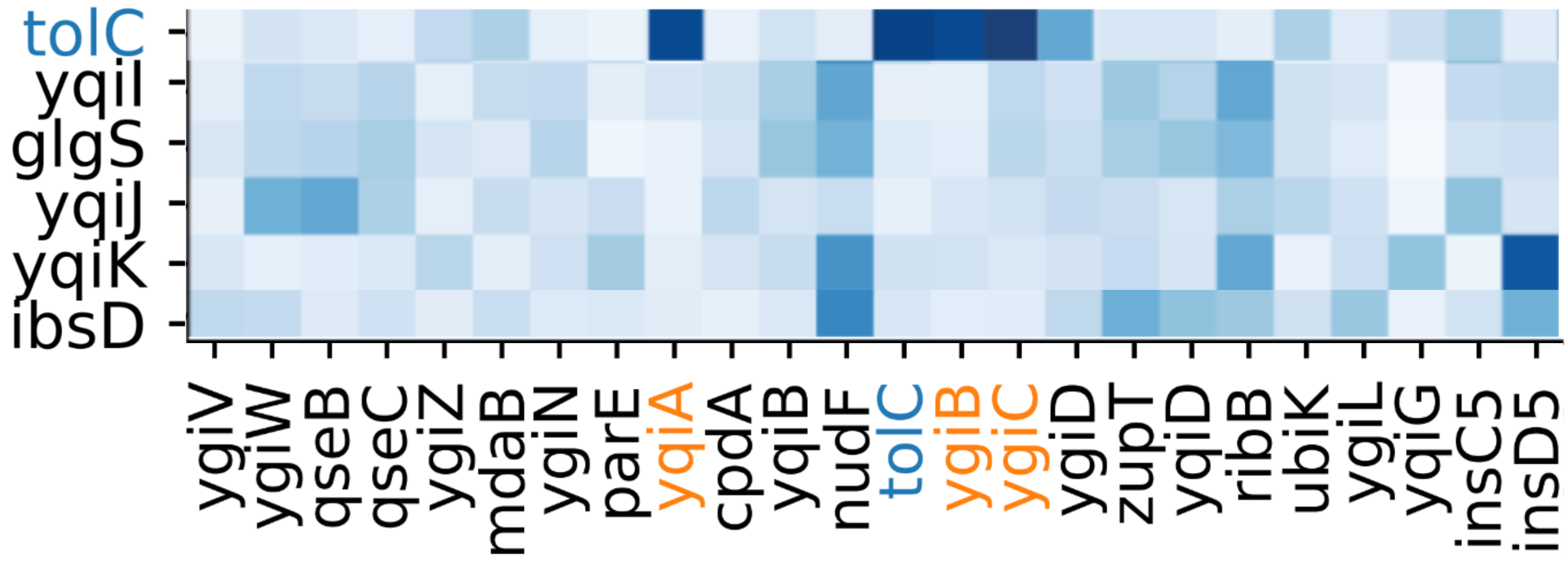}   
    \vspace{-1.5em}
    \caption{Visualization of Attention. 
        }
    \label{fig:gene-operon}
    \vspace{-1.7em}
\end{wraptable}

\textbf{Results Analysis.}
The attention heatmap in Fig.~\ref{fig:gene-operon} shows that gene operon \textcolor{mycolor2}{tolC} has high attention weight with the operons \textcolor{mycolor1}{ygiB} and \textcolor{mycolor1}{ygiC} and \textcolor{mycolor1}{yqiA}. This suggests a significant interaction among these operons, indicating the presence of a shared genetic operon tolC-ygib.  This inference finds support in biological research, which associates these operons with the DUF1190 domain-containing protein YgiB~\cite{karp2019biocyc}.

\textbf{Level 2 Task B: Functional Gene Prediction $\rightarrow $  Antimicrobial Resistance Genes (ARG) Prediction}.
This task is crucial for understanding ARGs and facilitates the identification of resistance mechanisms.
However, existing methods suffer from high false-positive rates and category bias~\cite{jian2021antibiotic, arnold2022horizontal}, necessitating the use of deep learning methods to rapidly and accurately detect ARG presence in metagenomic data.
The CARD dataset~\cite{jia2016card} categorizes each gene into one of 269 AMR Gene Families (CARD-A), 37 Drug Classes (CARD-D), or 7 Resistance Mechanisms (CARD-R). 
Our model performs a three-category classification for each gene sequence in CARD, respectively.

\begin{wrapfigure}{r}{0.5\textwidth}
    \centering
    \vspace{-1.5em}
    \centering    
    \captionof{table}{Classification Results on CARD-R.}
    \vspace{-0.5em}
    \label{tab:CARD_ours_Resistance_ratio}
    \scalebox{0.8}{
        \begin{tabular}{p{4.5cm}p{0.8cm}p{0.8cm}p{0.8cm}}
            \toprule
            \begin{tabular}[c]{@{}l@{}}\textbf{Class}\textbf{Name}\end{tabular} & \begin{tabular}[c]{@{}l@{}}\textbf{Total}\\\textbf{Count}\end{tabular} & \begin{tabular}[c]{@{}l@{}}\textbf{Correct}\\\textbf{Num}\end{tabular} & \begin{tabular}[c]{@{}l@{}}\textbf{Correct}\\\textbf{Ratio}\end{tabular} \\ 
            \midrule
            Antibiotic Inactivation                             & 252                                                  & 238                                                  & 94.44\%                                                \\
            Antibiotic Target Alteration                        & 70                                                   & 59                                                   & 84.29\%                                                \\
            Antibiotic Target Protection                        & 28                                                   & 27                                                   & 96.43\%                                                \\
            Antibiotic Efflux                                   & 25                                                   & 18                                                   & 72.00\%                                                \\
            Antibiotic Target Replacement                       & 14                                                   & 12                                                   & 85.72\%                                                \\ 
            \bottomrule
        \end{tabular}}        
    \vspace{-1.5em}
\end{wrapfigure}

\textbf{Results Analysis.}
{FGBERT}'s performance on CARD-A is significant, as shown in Tab.~\ref{tab:new-table}. 
This category's broad range (269 classifications) creates a long-tail distribution, necessitating an understanding of the biological properties of gene sequences for accurate annotation~\cite{mcarthur2013comprehensive}.
To mitigate this issue, we adjust the data sampling strategy to increase the frequency of fewer samples, improving the model's prediction accuracy.
Tab.~\ref{tab:CARD_ours_Resistance_ratio} demonstrates {FGBERT}'s high prediction accuracy for the CARD-R category, exhibiting superior classification results for both majority and minority classes, with over 85\% accuracy.
Appendix Tab.~\ref{tab:CARD_ours_AMR_ratio} reveals {FGBERT} has better performance for all 269 AMR Gene Family categories, with 100\% classification accuracy for majority categories like CTX, ADC, CMY, as well as for minor ones like AXC, CRH, KLUC.

\begin{table*}[t]
        \vspace{-2.5em}
        \centering
        \caption{Macro F1 ($\% \uparrow$) and Weighted F1 ($\% \uparrow$) on eight downstream tasks. This includes Gene Operon Prediction on E-K12, ARG Prediction on three CARD categories, Virulence Factors Classification on VFDB, Enzyme Function Annotation on ENZYME, Microbial Pathogens Detection on PATRIC, Nitrogen Cycle Processes Prediction on NCycDB. RF denotes Random Forest, and VT represents Vanilla Transformer.
        The highest results are highlighted with \textcolor{red}{\textbf{boldface}}. The second highest results are highlighted with \textcolor[rgb]{1,0.647,0}{\uline{underline}}.   
        }
        \vspace{-0.5em}
        \label{tab:new-table}
        \footnotesize
        \begin{adjustbox}{width=\textwidth}
                \begin{tabular}
                        {p{1.3cm}p{0.4cm}p{0.4cm}p{0.04mm}p{0.4cm}p{0.4cm}p{0.4cm}p{0.4cm}p{0.4cm}p{0.4cm}p{0.04mm}p{0.4cm}p{0.4cm}p{0.04mm}p{0.4cm}p{0.4cm}p{0.04mm}p{0.4cm}p{0.4cm}p{0.04mm}p{0.4cm}p{0.4cm}}
                        \toprule
                        \multirow{3}{*}{Method} & \multicolumn{2}{c}{Operons~} &               & \multicolumn{6}{c}{ARG Prediction} &                            & \multicolumn{2}{c}{Virus~} &               & \multicolumn{2}{c}{Enzyme} &               & \multicolumn{2}{c}{Pathogen} &  & \multicolumn{2}{c}{N-Cycle}                                                                                                                                                    \\
                        \cline{2-3}\cline{5-10}\cline{12-13}\cline{15-16}\cline{18-19}\cline{21-22}
                                                & \multicolumn{2}{c}{E-K12}    &               & \multicolumn{2}{c}{CARD-A}         & \multicolumn{2}{c}{CARD-D} & \multicolumn{2}{c}{CARD-R} &               & \multicolumn{2}{c}{VFDB}   &               & \multicolumn{2}{c}{ENZYME}   &  & \multicolumn{2}{c}{PATRIC}  &               & \multicolumn{2}{c}{NCycDB}                                                                                                       \\
                        \cline{2-3}\cline{5-10}\cline{12-13}\cline{15-16}\cline{18-19}\cline{21-22}
                                                & M.F1                         & W.F1          &                                    & M.F1                       & W.F1                       & M.F1          & W.F1                       & M.F1          & W.F1                         &  & M.F1                        & W.F1          &                            & M.F1          & W.F1          &  & M.F1          & W.F1          &  & M.F1          & W.F1          \\
                        \midrule
                        RF                      & 20.2                         & 34.8          &                                    & 22.4                       & 35.3                       & 36.1          & 49.0                       & 47.8          & 57.6                         &  & 22.4                        & 38.5          &                            & 33.6          & 41.2          &  & 25.3          & 29.8          &  & 67.0          & 71.7          \\
                        SVM                     & 38.6                         & 45.2          &                                    & 27.6                       & 40.5                       & 33.6          & 47.2                       & 43.3          & 66.2                         &  & 28.0                        & 41.4          &                            & 31.3          & 43.6          &  & 26.6          & 31.2          &  & 66.9          & 70.3          \\
                        KNN                     & 39.9                         & 41.0          &                                    & 36.9                       & 54.4                       & 36.4          & 51.3                       & 36.2          & 63.5                         &  & 27.3                        & 47.1          &                            & 31.4          & 42.9          &  & 11.0          & 27.4          &  & 68.8          & 73.2          \\
                        \midrule
                        LSTM                    & 40.4                         & 42.5          &                                    & 47.1                       & 60.3                       & 39.1          & 62.3                       & 47.5          & 84.2                         &  & 36.7                        & 66.3          &                            & 42.8          & 51.0          &  & 41.3          & 49.7          &  & 71.9          & 81.2          \\
                        BiLSTM                  & 38.2                         & 43.8          &                                    & 47.4                       & 61.9                       & 43.5          & 58.1                       & 58.9          & 80.3                         &  & 46.1                        & 72.1          &                            & 38.7          & 50.2          &  & 43.3          & 48.5          &  & 82.0          & 88.4          \\
                        VT                      & 43.3                         & 47.8          &                                    & 57.1                       & 70.0                       & 49.8          & 68.1                       & 55.7          & 86.4                         &  & 58.0                        & 81.0          &                            & 68.2          & 75.8          &  & 49.8          & 57.3          &  & 84.5          & 90.7          \\
                        \midrule
                        HyenaDNA                       & 42.4                                     & 47.1                                     &  & 50.9                                     & 68.2                                     & 53.6                                     & 78.1                                     & 66.2                                     & 88.1                                     &  & \textcolor[rgb]{1,0.647,0}{\uline{61.0}} & 70.4                                     &  & 79.6                                     & 83.6                                     &  & 51.1                                     & 57.6                                     &  & 92.4                                     & 96.0                                      \\
                        ESM-2                          & 38.2                                     & 42.5                                     &  & 57.2                                     & 71.4                                     & 56.0                                     & \textcolor[rgb]{1,0.647,0}{\uline{82.1}} & \textcolor[rgb]{1,0.647,0}{\uline{68.2}} & 90.0                                     &  & 60.7                                     & \textcolor[rgb]{1,0.647,0}{\uline{84.4}} &  & \textcolor[rgb]{1,0.647,0}{\uline{92.5}} & \textcolor[rgb]{1,0.647,0}{\uline{96.7}} &  & \textcolor[rgb]{1,0.647,0}{\uline{56.0}} & \textcolor[rgb]{1,0.647,0}{\uline{67.5}} &  & \textcolor[rgb]{1,0.647,0}{\uline{95.8}} & \textcolor[rgb]{1,0.647,0}{\uline{96.1}}  \\
                        NT                             & 45.1                                     & 44.8                                     &  & 58.5                                     & 72.0                                     & \textcolor[rgb]{1,0.647,0}{\uline{56.2}} & 80.2                                     & 68.0                                     & \textcolor[rgb]{1,0.647,0}{\uline{90.3}} &  & 58.3                                     & 71.6                                     &  & 74.1                                     & 76.7                                     &  & 46.1                                     & 61.9                                     &  & 75.1                                     & 86.5                                      \\
                        DNABERT2                       & \textcolor[rgb]{1,0.647,0}{\uline{51.7}} & \textcolor[rgb]{1,0.647,0}{\uline{52.4}} &  & \textcolor[rgb]{1,0.647,0}{\uline{65.2}} & \textcolor[rgb]{1,0.647,0}{\uline{79.8}} & 51.5                                     & 78.7                                     & 61.2                                     & 88.6                                     &  & 58.2                                     & 82.3                                     &  & 85.4                                     & 85.2                                     &  & 52.9                                     & 60.6                                     &  & 88.6                                     & 95.7                                      \\ 
                        \midrule
                        \textbf{\textcolor{red}{Ours}} & \textbf{\textcolor{red}{61.8}}           & \textbf{\textcolor{red}{65.4}}           &  & \textbf{\textcolor{red}{78.6}}           & \textbf{\textcolor{red}{90.1}}           & \textbf{\textcolor{red}{57.4}}           & \textbf{\textcolor{red}{85.2}}           & \textbf{\textcolor{red}{69.4}}           & \textbf{\textcolor{red}{91.4}}           &  & \textbf{\textcolor{red}{75.7}}           & \textbf{\textcolor{red}{90.2}}           &  & \textbf{\textcolor{red}{99.1}}           & \textbf{\textcolor{red}{98.8}}           &  & \textbf{\textcolor{red}{99.3}}           & \textbf{\textcolor{red}{99.0}}           &  & \textbf{\textcolor{red}{99.5}}           & \textbf{\textcolor{red}{99.2}}            \\
                        \toprule
                \end{tabular}
        \end{adjustbox}
        \vspace{-2.5em}
\end{table*}

\textbf{Level 2 Task C: Functional Gene Prediction $\rightarrow$  Virulence Factors (VF) Prediction.} This task is to detect microbial elements like bacterial toxins, which enhance pathogen infectivity and exacerbate antimicrobial resistance.
Existing methods for metagenomic analysis, particularly those co-predicting ARGs and VFs, are inadequate and suffer from threshold sensitivity issues~\cite{yang2016args}. 
VFDB dataset~\cite{chen2005vfdb} includes the major VFs of the most characterized bacterial pathogens, detailing their structural features, functions, and mechanisms. We use VFDB core dataset, including 8945 VF sequences and 15 VF categories.
\textbf{Results Analysis.}
Our model achieves the SOTA results on VFDB, as reported in Tab.~\ref{tab:new-table}. 
It significantly outperforms the genomic pre-trained model; for instance, M.F1 and W.F1 scores improve by 30\% and 9.5\%, respectively, compared to DNABERT2. 
This highlights the limitations of directly applying the genomic pre-trained model to metagenomic data for precise functional annotation. Conversely, ESM-2, as a PLM, excels by leveraging intrinsic protein information in metagenomic data, highlighting its effectiveness.

\textbf{Level 2 Task D: Functional Gene Prediction $\rightarrow$  Enzyme Function Prediction}. This task is critical for understanding metabolism and disease mechanisms in organisms. While traditional methods rely on time-consuming and labor-intensive biochemical experiments, advanced technologies can offer efficient and accurate predictions for large-scale genomic data. ENZYME~\cite{bairoch2000enzyme} is a repository of information related to the nomenclature of enzymes.
We organize 5761 data, 7 categories of ENZYME core dataset, each enzyme has a unique EC number~\cite{nomenclature1992recommendations}.
\textbf{Results Analysis.}
Our experimental results demonstrate {FGBERT}'s superior performance on the ENZYME dataset. It outperforms ESM-2, the second-highest method, by approximately 6.62\% in M.F1 and 2.09\% in W.F1, demonstrating its ability to discern distinct enzyme function characteristics. This observation highlights that our model not only captures gene-protein contextual relationships but also effectively models the relationships between sequences and functions within metagenomic data.

\textbf{Level 3 Task E: Pathogenicity Potential Assessment $\rightarrow $  Genome Pathogens Prediction.} This task assesses the pathogenic potential of pathogens to cope with the public health risks caused by newly emerging pathogens.
Accurate deep-learning algorithms are key for the precise identification of pathogens, improving the ability to respond to drug resistance threats. 
We use PATRIC core dataset~\cite{gillespie2011patric}, which has 5000 pathogenic bacterial sequences across 110 classes.
\textbf{Results Analysis.} Tab.~\ref{tab:new-table} shows {FGBERT}'s classification of pathogenic bacteria species within PATRIC, demonstrating superior performance over baselines by recognizing crucial genera features.
The PATRIC dataset presents a significant challenge due to its large number of categories and sparse data.
Baselines generally underperform because they require more data to discern the subtle differences between numerous categories. In contrast, {FGBERT} stands out with M.F1 and W.F1 scores of 99.27\% and 99.03\%, respectively.
This robust performance indicates its advanced learning capability, making it well-suited for high-dimensional classification tasks and highlighting the benefits of using protein-based gene representations for enhanced functional annotation accuracy.

\textbf{Level 4 Task F:  Nitrogen Cycle Prediction $\rightarrow $  Nitrogen (N) Cycling Process Prediction.} This task focuses on the functional genes related to the N cycle, linking them to environmental and ecological processes.
NCycDB~\cite{tu2019ncycdb} contains 68 gene (sub)families and covers 8 N cycle processes with 213,501 representative sequences at 100\% identity cutoffs, each involving a specific gene family.
\textbf{Results Analysis.} Tab.~\ref{tab:new-table} presents {FGBERT}'s classification results on NCycDB, suggesting its ability to recognize key features of particular N cycle processes and improve gene family classification by recognizing domains.
Although baselines show improved performance on NCycDB compared to PATRIC, due to a larger amount of data per category aiding in discrimination among diverse categories, {FGBERT} still leads with macro F1 (M.F1) and weighted F1 (W.F1) scores of 99.49\% and 99.22\%, respectively. 
However, pre-trained baselines require more time and memory for tokenising large datasets, as analyzed in Sec.~\ref{model_efficiency}.

\subsection{Ablation Study}

\begin{wrapfigure}{r}{0.5\textwidth}
    \centering
    \begin{minipage}{0.49\textwidth}
        \vspace{-3.8em}
        \captionof{table}{Ablation Study of M.F1 on CARD-D. 
         }
        \label{new-ablation-study}
        \vspace{-0.5em}
        \resizebox{\linewidth}{!}{
            \begin{tabular}{llllll}
                \toprule
                \multirow{2}{*}{Method} & Operons                           &  & \multicolumn{3}{c}{ARG Prediction}                                                                         \\
                \cline{2-2}\cline{4-6}
                                        & E-K12                             &  & CARD-A                             & CARD-D                            & CARD-R                            \\
                \midrule
                \textbf{{FGBERT}}        & \multicolumn{1}{c}{\textbf{61.8}} &  & \multicolumn{1}{c}{\textbf{78.7}}  & \multicolumn{1}{c}{\textbf{57.4}} & \multicolumn{1}{c}{\textbf{69.4}}
                \\
                $w/o.$ MGM              & \multicolumn{1}{c}{-8.1}          &  & \multicolumn{1}{c}{-6.7}           & \multicolumn{1}{c}{-10.5}         & \multicolumn{1}{c}{-6.7}          \\
                $w/o.$ Triplet          & \multicolumn{1}{c}{-7.4}          &  & \multicolumn{1}{c}{-5.4}           & \multicolumn{1}{c}{-5.8}          & \multicolumn{1}{c}{-3.4}
                \\
                \bottomrule
                \end{tabular}}  
    \vspace{0.5em}
    \end{minipage}
    \begin{minipage}{0.49\textwidth}
        \centering
        \includegraphics[width=0.9\textwidth]{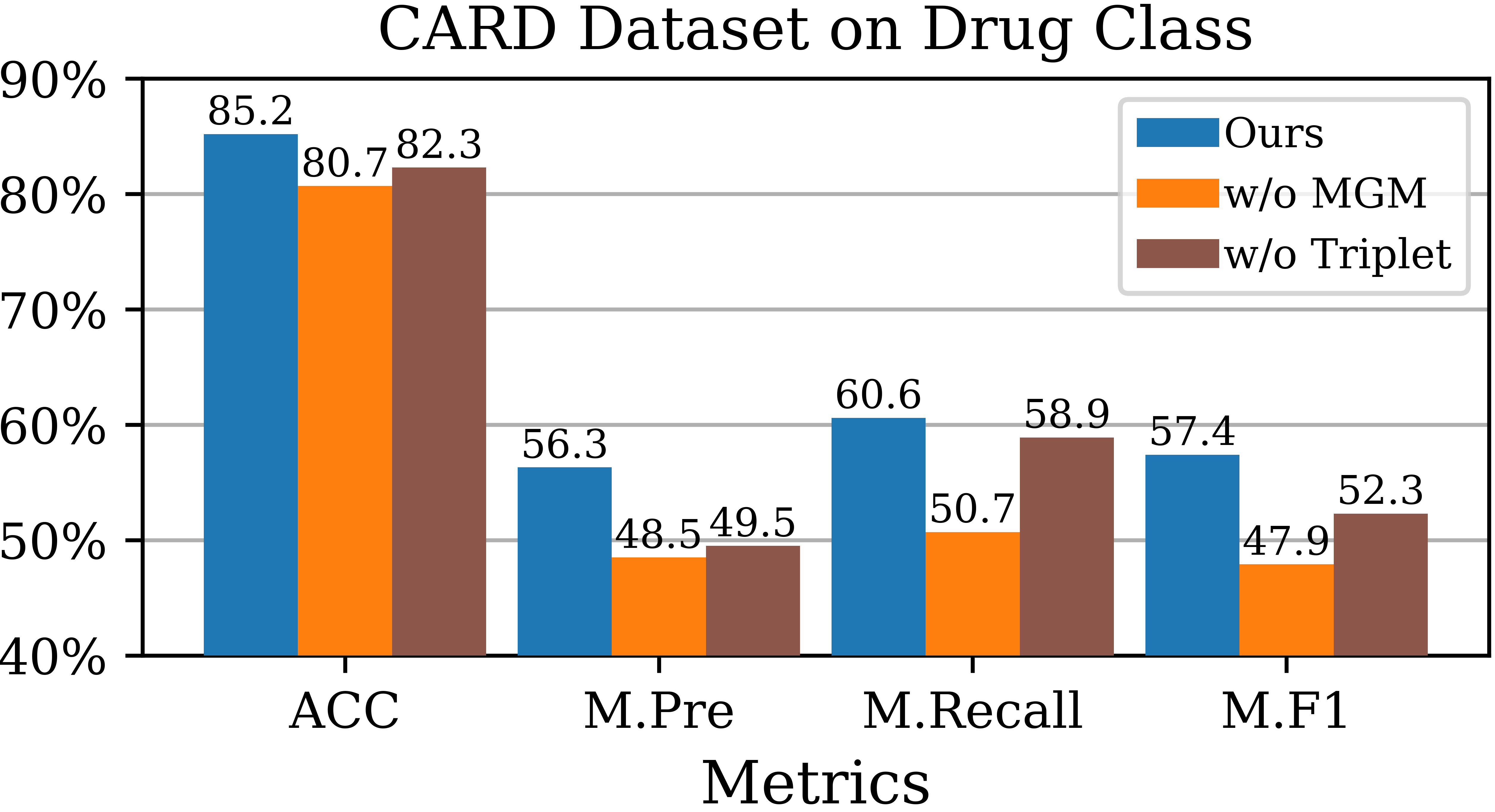}
        \vspace{-0.5em}
        \caption{Ablation studies of Our Proposed Modules on Four Downstream Tasks.}
        \label{fig:ablation-study}
        \vspace{-2.1em}
    \end{minipage}    
\end{wrapfigure}

\textbf{Ablation Study on the Performance of MGM and TMC.} We perform the ablation study to investigate the effectiveness of our proposed components.
Tab.~\ref{new-ablation-study} compares the performance of {FGBERT} without MGM and TMC module across four datasets. 
The decrease in M.F1 score after removing the MGM and TMC, respectively, highlights their roles in enhancing model performance across different tasks.
While TMC contributes to performance, it is evident that MGM has a more substantial impact.
Fig.~\ref{fig:ablation-study} illustrates four more metrics on the CARD dataset.

\textbf{Ablation Study on Visualization Results of MGM.} We conduct a visualization experiment to validate the effectiveness of MGM in the One-to-Many scenario.
ATP synthases can exhibit different functions in different organisms to adapt to their respective environmental conditions, even though the basic functions are the same~\cite{hong2008atp}.

\begin{wraptable}{r}{0.6\textwidth}
    \centering
    \vspace{-1.5em}
    \hspace{-0.1cm}
    \includegraphics[height=0.40\linewidth,keepaspectratio]{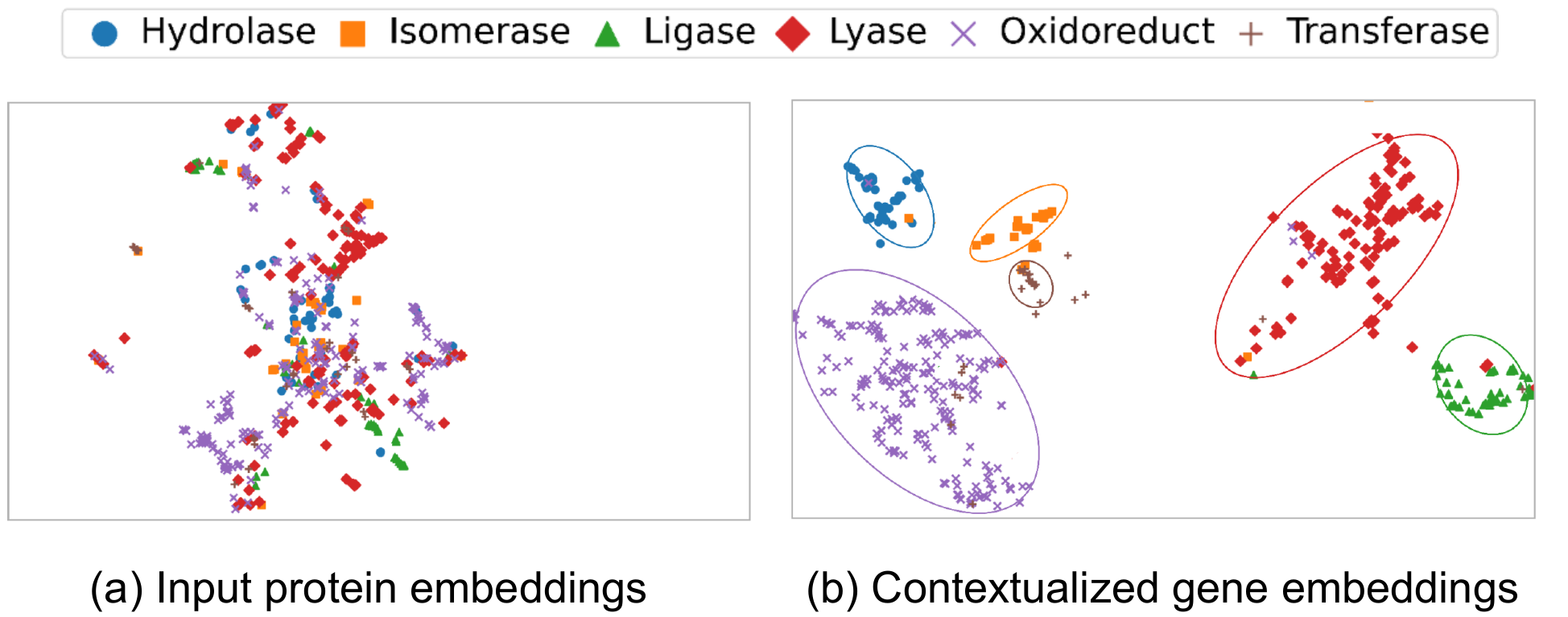}
    \vspace{-0.5em}
    \caption{T-SNE Visualization of Different Embeddings for ATP Synthases. Each dot denotes a sequence and is colored according to different functions.}
    \label{fig:t-sne-plot}
    \vspace{-1.4em}
\end{wraptable}

We collect 1177 ATP synthase sequences from UniProt~\cite{uniprot2023uniprot} and color them according to six taxonomies, i.e. Hydrolases (35.6\%, dark green), Isomerases (3.6\%, orange), Ligases (35.7\%, blue), Lyases (10.7\%, pink), Oxidoreduct (8.9\%, light green) and Transferases (5.3\%, yellow) (numbers in parentheses indicate category ratios). 
Fig.\ref{fig:t-sne-plot} (a) shows the clustering results of ATPase protein embeddings without genome contextual analysis, indicating that more dispersed clustering results of genes in different genome contexts.
Fig.\ref{fig:t-sne-plot} (b) presents the clustering results of ATPase embeddings after our designed protein-based gene embedding, which shows genes belonging to the same category (the same color) apparently cluster together.
Concretely, Isomerases (3.5\%, orange) and Transferases (5.3\%, yellow), which account for the smallest percentage, cluster together, whereas, in the left plot, these two are scattered. This demonstrates that our proposed MGM can resolve the One-to-Many problem between sequences and functions.


\begin{wrapfigure}{r}{0.4\textwidth}
    \centering
    \vspace{-1.3em}
    \captionof{table}{Clustering Results of TMC for Many-to-One problem.}
    \vspace{-0.5em}
    \label{tab:cluster-many-to-one}
    \resizebox{1\linewidth}{!}{ \begin{tabular}{llll}
        \toprule
        Method     & NMI  & ARI  & \begin{tabular}[c]{@{}l@{}}Silhouette\\Coefficient\end{tabular} \\
        \midrule
        +Tokenizer & 0.44 & 0.34 & 0.7                                                             \\
        +MGM       & 0.66 & 0.51 & 0.72                                                            \\
        +TMC       & \textbf{0.72} & \textbf{0.59} & \textbf{0.75}                                                            \\
        \bottomrule
        \end{tabular}}
    \vspace{-1.8em}
\end{wrapfigure}

\textbf{Ablation Study on Clustering results of TMC.} We further explore the impact of integrating TMC into our model on gene operon prediction. Initially, we evaluate {FGBERT} without TMC module. Afterwards, we add TMC module, and K-means are performed using the embeddings from the network's intermediate layers. We compute the Normalized Mutual Information (NMI), Adjusted Rand Index (ARI), and Silhouette Coefficient Index to measure the clustering quality.
The results in Tab.~\ref{tab:cluster-many-to-one} show significant improvements in ARI and Silhouette Coefficient with TMC, highlighting its effectiveness in enhancing clustering performance across various metrics.

\textbf{Ablation Study on Context-Aware Tokenizer.}
We replace the {FGBERT}'s tokenizer with ESM2 and BPE representations and assessed performance across downstream tasks. Tab.~\ref{tab:ablation-tokenizer} shows that {FGBERT} outperforms both methods in capturing complex gene sequence-function relationships. Although DNABERT2 compresses sequences effectively with BPE, our context-aware tokenizer demonstrates superior accuracy in metagenomic data analysis. For complete results, see Appendix Tab.~\ref{tab:app-ablation-tokenizer}.

\vspace{-1em}
\begin{table}[t]
\centering
\caption{The ablation study using protein-based gene representation as a context-aware tokenizer.}
\label{tab:ablation-tokenizer}
\footnotesize
\begin{tabular}{ll|lllllll} 
\toprule
Dataset & Method & Acc                                      & M.Pre                                    & M.Recall                                 & M.F1                                     & W.Pre                                    & W.Recall                                 & W.F1                                      \\ 
\midrule
E-K12   & FGBERT & \textbf{\textcolor{red}{0.68}}           & \textbf{\textcolor{red}{0.68}}           & \textbf{\textcolor{red}{0.61}}           & \textbf{\textcolor{red}{0.61}}           & \textbf{\textcolor{red}{0.77}}           & \textbf{\textcolor{red}{0.67}}           & \textbf{\textcolor{red}{0.65}}            \\
        & ESM2   & 0.54                                     & 0.51                                     & 0.48                                     & 0.49                                     & 0.61                                     & 0.51                                     & 0.52                                      \\
        & BPE    & \textcolor[rgb]{1,0.647,0}{\uline{0.59}} & \textcolor[rgb]{1,0.647,0}{\uline{0.58}} & \textcolor[rgb]{1,0.647,0}{\uline{0.55}} & \textcolor[rgb]{1,0.647,0}{\uline{0.54}} & \textcolor[rgb]{1,0.647,0}{\uline{0.67}} & \textcolor[rgb]{1,0.647,0}{\uline{0.54}} & \textcolor[rgb]{1,0.647,0}{\uline{0.55}}  \\ 
\hline
CARD-A  & FGBERT & \textbf{\textcolor{red}{0.91}}           & \textbf{\textcolor{red}{0.77}}           & \textbf{\textcolor{red}{0.8}}            & \textbf{\textcolor{red}{0.78}}           & \textbf{\textcolor{red}{0.9}}            & \textbf{\textcolor{red}{0.91}}           & \textbf{\textcolor{red}{0.9}}             \\
        & ESM2   & 0.82                                     & 0.74                                     & 0.76                                     & 0.73                                     & 0.87                                     & 0.82                                     & 0.81                                      \\
        & BPE    & \textcolor[rgb]{1,0.647,0}{\uline{0.84}} & \textcolor[rgb]{1,0.647,0}{\uline{0.75}} & \textcolor[rgb]{1,0.647,0}{\uline{0.77}} & \textcolor[rgb]{1,0.647,0}{\uline{0.75}} & \textcolor[rgb]{1,0.647,0}{\uline{0.88}} & \textcolor[rgb]{1,0.647,0}{\uline{0.86}} & \textcolor[rgb]{1,0.647,0}{\uline{0.86}}  \\ 
\bottomrule
\end{tabular}
\vspace{-6mm}
\end{table}

\subsection{Model Efficiency Analysis}\label{model_efficiency}


\begin{wraptable}{r}{0.45\textwidth}
    \centering
    \vspace{-3.8em}
    \hspace{-0.1cm}
    \includegraphics[height=0.5\linewidth,keepaspectratio]{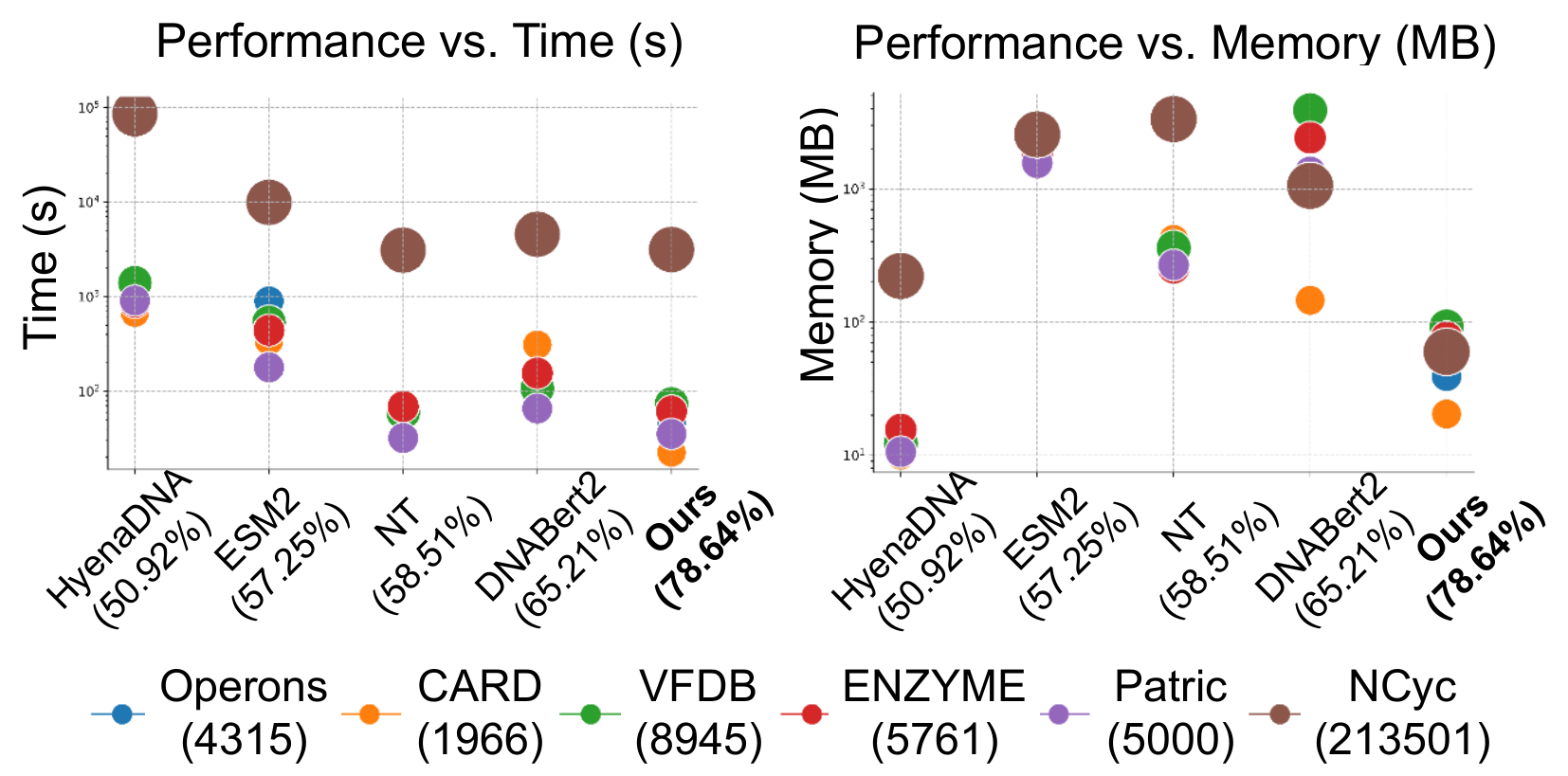}
    \vspace{-1.5em}
    \caption{Comparative Analysis on Tokenization Efficiency: Time(s) vs. Memory (MB). Each point denotes a specific dataset, with the size indicating its scale.}
    \label{fig:Token-Analysis}
    \vspace{-4.em}
\end{wraptable}

We analyze the time complexity and memory efficiency of our tokenizer compared to four genomic pre-trained methods on six datasets in Fig.~\ref{fig:Token-Analysis}. Our tokenizer demonstrates superior efficiency, achieving a significant reduction in both time and memory usage.
Notably, on NCyc dataset (brown) with 213,501 sequences, ours reduces processing time by 31.05\% and memory usage by 94.33\% compared to DNABERT2.
For the CARD dataset (orange) with 1,966 sequences, time and memory usage decreased by 61.70\% and 58.53\%.
Although HyenaDNA uses less memory on Operons, CARD, VFDB, and ENZYME datasets, it underperforms ours in time cost and overall performance. 

\begin{table}[t]
\centering
\vspace{-1em}
\caption{The Scalability of FGBERT.}
\label{tab:time-memory-efficiency}
\begin{adjustbox}{width=1\textwidth}
\begin{tabular}{lllllllll}
    \toprule
    \Huge
    Model    & \Huge \makecell{Pre-training\\ Data} & \Huge \makecell{\# Layers\\} & \Huge \makecell{Hidden\\ Dim} & \Huge \makecell{\# Heads\\ } & \Huge \makecell{Operons \\(500/1000/1500 ep) }& \Huge \makecell{CARD-A \\(500/1000/1500 ep)} & \Huge \makecell{CARD-D\\ (500/1000/1500 ep)} & \Huge \makecell{CARD-R \\(500/1000/1500 ep)} \\     
    \midrule
    \Huge FGBERT-T &\Huge  50M               &\Huge  10     &\Huge  640        &\Huge  5     &\Huge  50.9 / 61.5 / 63.5         &\Huge  74.6 / 88.6 / 90.7       &\Huge  72.4 / 83.7 / 86.4        &\Huge  77.3 / 90.1 / 91.1        \\
    \Huge FGBERT-S &\Huge  100M              &\Huge  19     &\Huge  1280       &\Huge  10    &\Huge  55.1 / 65.4 / 65.9         &\Huge  80.4 / 90.1 / 91.2        &\Huge  76.9 / 85.2 / 87.5        &\Huge  81.4 / 91.4 / 93.0          \\
    \Huge FGBERT-B &\Huge  150M              &\Huge  25     &\Huge  2560       &\Huge  25    &\Huge  57.1 / 66.7 / 67.6         &\Huge  82.7 / 91.1 / 93.2        &\Huge  81 / 87.6 / 90.0            &\Huge  83.7 / 93.4 / 94.8        \\
    \bottomrule
\end{tabular}
\end{adjustbox}
\vspace{-0.5em}
\end{table}
To further explore model scalability, we train three FGBERT variants—FGBERT-T, FGBERT-S, and FGBERT-B—differing in layers, hidden dimensions, and attention heads. Each variant is trained on 50 million, 100 million, and 150 million sequences to assess the impact on time and memory during pre-training, as shown in Tab.~\ref{tab:time-memory-efficiency}.

\subsection{Sensitivity Analysis}

\begin{wraptable}{r}{0.6\textwidth}
    \centering
    \vspace{-1.5em}
    \hspace{-0.1cm}
    \includegraphics[width=\linewidth]{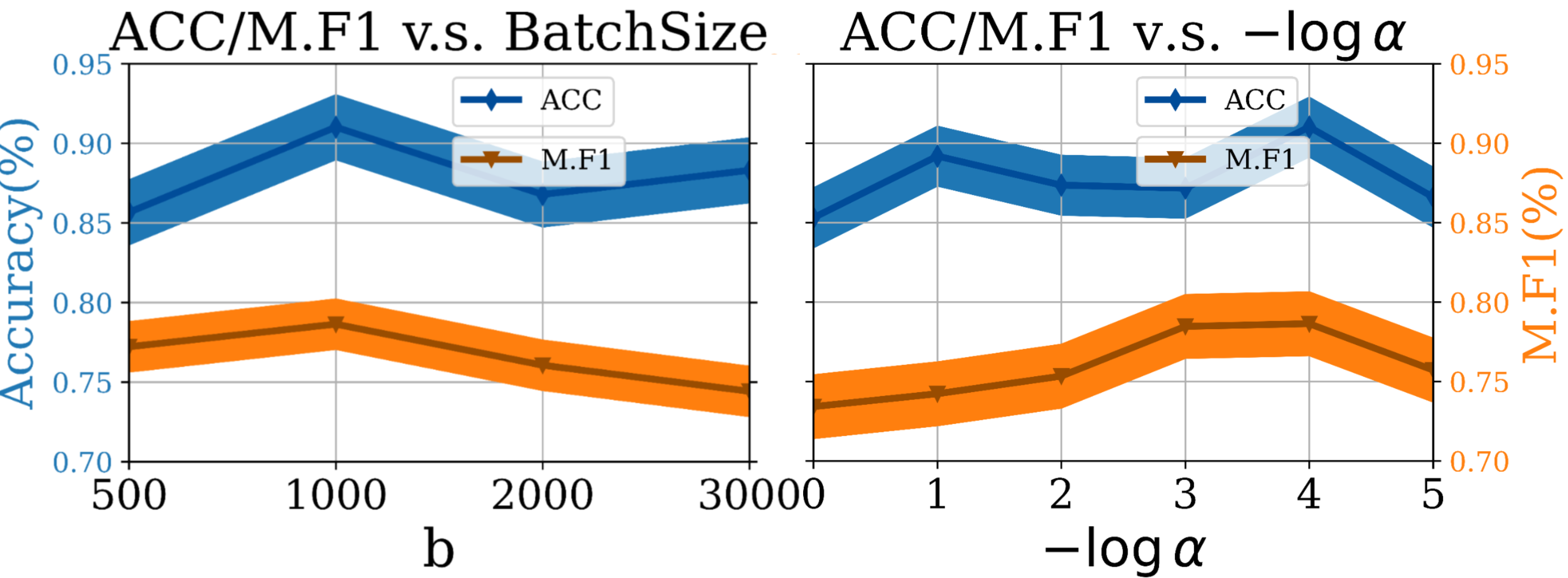}
    \caption{Sensitivity w.r.t Hyper-parameters $\alpha, b$ of CARD dataset on AMR Gene Family.}
    \label{fig:Sensitivity Analysis}
    \vspace{-1.4em}
\end{wraptable}

Our sensitivity analysis indicates that {FGBERT} can be optimized effectively using small batch size $b$ without larger performance degradation shown in Fig.~\ref{fig:Sensitivity Analysis}.
This is important for resource-constrained scenarios, highlighting our model maintains good performance even with limited data.
We choose a batch size of 1000.
Next, we analyze another hyper-parameter of balance ratio $\alpha$ on CARD dataset. {FGBERT} obtains good performance for different values of $\alpha$, demonstrating its robustness and insensitivity to this hyper-parameter. We set $\alpha$ to be 0.4.

\subsection{Model Configurations.}
All our experiments are performed on 4 NVIDIA V100 GPUs and the PyTorch framework. 
The encoder of {FGBERT} is initialized with Roberta~\cite{devlin2018bert}. 
For pre-training, the parameters are set as follows: batch size of 1000, 19 encoder layers, 10 attention heads, an embedding dimension of 1280, and relative position encoding. During the pre-training stage, the model is trained for 500 epochs using the AdamW optimizer~\cite{loshchilov2017decoupled} with a weight decay of 0.02. The learning rate starts at 1e-5, warms up to 1e-4 over the first 5000 steps, and then decreases back to 1e-5 following a cosine decay schedule.
The complete model consists of 954.73 million parameters and requires 2.55 billion FLOPs, as detailed in Appendix Tab.~\ref{tab:parameters}.

%% file: sec_conclusion.tex
In this paper, we propose a new idea of protein-based gene representation, preserving essential biological characteristics within each gene sequence. With the new context-aware tokenizer, we propose MGM, a gene group-level pre-training task, designed to learn the interactions between genes. Additionally, we develop TMC, a contrastive learning module to generate multiple positive and negative samples to distinguish the gene sequences. MGM and TMC constitute a joint pre-training model, {FGBERT} for metagenomic data.
Our experiments and visualizations demonstrate the superior performance of our model.
For the future, it remains to be explored how to incorporate multi-omics data, such as metabolomics, into our metagenomic pre-trained model.

%% file: sec_appendix.tex
\appendix

\section{Appendix / supplemental material}
Optionally include supplemental material (complete proofs, additional experiments and plots) in appendix.
All such materials \textbf{SHOULD be included in the main submission.}

\section{Methods.}
\subsection{Negative Sample Sampling Strategy.}
The selection of negative samples is mapped based on the Euclidean distance between EC numbers. Each EC number is considered a unique functional category or cluster, with seven clusters corresponding to seven distinct EC numbers.
For each anchor sample, ten negative samples are chosen based on their proximity within the distance mapping, selecting the nearest in terms of Euclidean distance.
To provide a concise overview, the clusters and their corresponding entries are summarized in Tab.~\ref{tab:negative-samlpe}.
\begin{table}[ht]
\centering
\caption{The specific details of the negative sample sampling strategy.}
\label{tab:negative-samlpe}
\begin{tabular}{lll} 
\hline
EC & Entry  & EC number  \\ 
\hline
1  & F2JXJ3 & 1.4.3.20   \\
2  & Q3TTA7 & 2.3.2.27   \\
3  & Q8K337 & 3.1.3.36   \\
4  & Q9LZI2 & 4.1.1.35   \\
5  & Q79FW5 & 5.4.2.12   \\
6  & Q9HN83 & 6.1.1.2    \\
7  & Q33821 & 7.1.1.2    \\
\hline
\end{tabular}
\end{table}

\section{Implementation Details.}
\subsection{Dataset.}
\textbf{MGnify dataset} includes genome sequence information from different microbial communities and their Biome Names (e.g., "Engineered", "Host-associated", and "Control"), as well as their corresponding Sample Classes, Sample Numbers, and Spectrum Categories. 
Lineage Category refers to the taxonomic information of microorganisms, which describes the taxonomic hierarchy of microorganisms from Domain (e.g., ”root” or ”Engineered”) to Species (e.g., ”Continuous culture” or ”Saliva”), which can be observed in Tab.~\ref{tab:MGnify_catalogues}.

\begin{table}[ht]
\centering
\caption{Microbial Species and Genomic Data Across Various Environments in the MGnify Dataset. Each column is catalogues, species, and genomes, respectively.}
\label{tab:MGnify_catalogues}
\begin{tabular}{lll}
\hline
Catalogues & Species & Genomes \\ \hline
Human gut & 4744 & 289232 \\
Human oral & 452 & 1225 \\
Cow rumen & 2729 & 5578 \\
Marine & 1496 & 1504 \\
Pig gut & 1376 & 3972 \\
Zebrafish feacel & 79 & 101 \\
Non-model fish gut & 172 & 196 \\ \hline
\end{tabular}
\end{table}

\begin{table}[htbp]
\centering
\begin{threeparttable}
\caption{Examples of Operons Dataset.}
\label{tab:operons_dataset}
\footnotesize 
\begin{tabularx}{\textwidth}{llXXX} 
\toprule
GeneID & Operon name & Description of Operon name                    & \begin{tabular}[c]{@{}l@{}}Gene name(s) \\ contained in the operon\end{tabular} & \begin{tabular}[c]{@{}l@{}}Other database's id\\ related to gene\end{tabular}  \\ 
\midrule
2973   & qseB        & DNA-binding transcriptional activator QseB & qseBC                                                                           & None                                                                           \\
2980   & cpdA        & cAMP phosphodiesterase                     & nudF-yqiB-cpdA-yqiA-parE                                                        & None                                                                           \\
2985   & ygiC        & putative acid–amine ligase YgiC            & tolC-ygiBC                                                                      & tolC-ygiBC                                                                     \\
\bottomrule
\end{tabularx}
\begin{tablenotes}
    \item Note: This table provides a partial list of instances from the Operons dataset, including Gene data IDs, operon names, descriptions, Gene name(s) contained in the operon, and other database's id related to the gene.
\end{tablenotes}
\end{threeparttable}
\end{table}

\addtocounter{table}{-1}

\begin{table}[htbp]
\centering
\caption{Examples of CARD Dataset}
\label{tab:CARD}
\footnotesize 
\begin{tabularx}{\textwidth}{llXXX} 
\toprule
Model ID & ARO Accession & AMR Gene Family & Drug Class & Resistance Mechanism \\ 
\midrule
1 & 3002999 & CblA beta-lactamase & cephalosporin & antibiotic inactivation \\
5 & 3002867 & trimethoprim resistant dihydrofolate reductase dfr & diaminopyrimidine antibiotic & antibiotic target replacement \\
7 & 3001989 & CTX-M beta-lactamase & cephalosporin & antibiotic inactivation \\
10 & 3002244 & CARB beta-lactamase & penam & antibiotic inactivation \\
20 & 3002012 & CMY beta-lactamase & cephamycin & antibiotic inactivation \\
\bottomrule
\end{tabularx}

\begin{tablenotes}
    \item Note: This table provides a partial list of instances from the CARD dataset, including data IDs, ARO (Antibiotic Resistance Ontology) Accession IDs, AMR (Antimicrobial Resistance) category names, drug category names, and Resistance Mechanism names.
\end{tablenotes}
\end{table}

\addtocounter{table}{-1}

\textbf{E.coil K12 RegulonDB}
Tab.~\ref{tab:operons_dataset} lists information such as operon names, descriptions, and names of genes included for some of the EK12 datasets.

\textbf{CARD}
Tab.~\ref{tab:CARD} shows a partial list of instances from the CARD dataset, including data IDs, ARO (Antibiotic Resistance Ontology) Accession IDs, AMR (Antimicrobial Resistance) category names, drug category names, and Resistance Mechanism names.

\subsection{Baselines}
We evaluate ours compared with three traditional machine learning methods (RandomForest, SVM, KNN), three sequence modeling methods (BiLSTM~\cite{schuster1997bidirectional} vs. LSTM~\cite{hochreiter1997long}), Vanilla Transformer and four pre-trained models on protein/DNA sequences(ESM-2~\cite{lin2023evolutionary}, HyenaDNA~\cite{nguyen2023hyenadna}, NT~\cite{dalla2023nucleotide} and DNABERT2~\cite{zhou2023dnabert}) for comparison.
Machine learning methods encode sequences by K-mers, where K takes the best one of the [3,4,5,6] parameters. Sequence modeling methods encode genes using Word2vec.
For a fair comparison, each model was trained and optimized to obtain the corresponding optimal hyperparameters.
We adopt a 5-fold cross-validation approach to partition all downstream task datasets into a training set (80\%), a validation set (10\%), and a test set (10\%).

\subsection{Model Configurations.}
All our experiments are performed on 4 NVIDIA V100 GPUs and the PyTorch framework. 
The encoder of {FGBERT} is initialized with Roberta~\cite{devlin2018bert}. 
The parameters for pre-training: batch size = 1000, 19 encoder layers, 10 attention heads, embedding dimension = 1280, and relative position encoding.
During the pre-training stage, the model is trained for 500 epochs. We use the AdamW optimizer~\cite{loshchilov2017decoupled} with a weight decay of 0.02. The learning rate is initialized to 1e-5 and warmed up to 1e-4 after 5000 epochs. Then, it is reduced to 1e-5 by a cosine decay strategy. 
The overall model comprises 954.73M parameters with a computational load of 2.55B FLOPs, detailed in Tab.~\ref{tab:parameters}.

\begin{table}[t]
\centering
\caption{The parameters and network layers learned by pre-training.}
\label{tab:parameters}
\begin{tabular}{ll} 
\toprule
\multicolumn{1}{l}{Batch
  Size} & 1000                        \\ 
\hline
Encoder
  Layer                   & 19                          \\
Attention
  Heads                 & 10                          \\
Embedding
  Dimension             & 1280                        \\
Positive
  Encoding               & Relative Positive Encoding  \\
Epochs                            & 1000                        \\
Optimizer                         & AdamW                       \\
Weight
  Decay                    & 0.02                        \\
Learning
  Rate                   & 1e-5→1e-4                   \\
Learning
  Rate Schedule          & Cosine                      \\
Warm-up
  Steps                   & 5000                        \\
\bottomrule
\end{tabular}
\end{table}

\subsection{Overview of Downstream Tasks} \label{app:overview of downstream}
\textbf{Operon Prediction Task} is to identify the transcription factor binding sites that have the strongest correlation with operon regulation in the gene regulatory network~\cite{cui2007micrornas, ermolaeva2001prediction}, which helps us to understand the mechanism and network of gene regulation and reveals the key interactions between transcription factors and operons~\cite{park2020enhancing}. Operons are generally composed of a set of multiple genes close to each other.

\textbf{ARGs and VFs Prediction Task.}
Pathogenic microorganisms pose a threat to public health by invading hosts through virulence factors (VFs) and exacerbating antibiotic resistance genes (ARGs)~\cite{o2016tackling}. 
Despite different evolutionary pathways, VFs and ARGs share common features that are critical for pathogenic bacteria to adapt to and survive in the microbial environment~\cite{martinez2002interactions}. 
Therefore, accurate identification of VFs and ARGs is extremely important for understanding the relationship between the microbiome and disease. 
However, traditional sequence-matching-based ARG identification methods suffer from high false-positive rates and specific ARG category bias~\cite{jian2021antibiotic, arnold2022horizontal}. 
Existing methods for analyzing metagenomic data are insufficient, especially tools for co-predicting the two are scarce, and also suffer from threshold sensitivity~\cite{yang2016args}.
Therefore, there is an urgent need for deep learning methods to rapidly and accurately predict the presence of VF and ARG in metagenomic datasets.

\textbf{Overview of Enzymes.} 
Enzymes are important catalysts in living cells that produce essential molecules needed by living organisms through chemical reactions~\cite{tawfik2010enzyme}. While traditional methods rely on time-consuming and labor-intensive biochemical experiments, advanced technologies such as deep learning can be used to efficiently and accurately predict large-scale genomic data, accelerating biomedical research and new drug development.

\textbf{Overview of Pathogens.} 
Pathogenic bacteria are a group of bacteria that can cause disease and can pose a significant threat to human health compared to commensal non-pathogenic bacteria in the human body with their ability to invade the host and cause disease~\cite{jones2008global}. 
The existing methods are limited to long genome sequences and are dominated by short sequence (<3 kb) inputs, which cannot process long genome sequences at the million bp level directly. Hence, the development of accurate deep learning algorithms for the precise identification of pathogens (causative agents) can help in the development of new therapies and vaccines and improve the ability to respond to the threat of drug resistance.

\textbf{Overview.} The Nitrogen cycle is a collection of important biogeochemical pathways in the Earth's ecosystems, and quantitatively studying the functional genes related to the N cycle and linking them to environmental and ecological processes is one of the important focuses in environmental genomics research~\cite{gruber2008earth}.
Currently, metagenome sequencing has been widely used to characterize gene families involved in N cycling processes. However, existing methods usually face the problems of inefficient database searches, unclear direct lineage classification, and low coverage of N cycle genes and/or gene (sub)families when analyzing N cycle gene families in metagenomes.
Consequently, accurate prediction of the functional categories of genes or proteins related to nitrogen metabolism in microorganisms is important for analyzing important resources in the process of nitrogen cycling and understanding the nitrogen cycle in ecosystems.

\section{Experimental Results.}


         
\subsection{More Detailed Ablation Experiments}
The results of Tab.~\ref{tab:more-abla} demonstrate each module's significant role in the overall performance.
\begin{table}[ht]
\centering
\caption{The result of {FGBERT} across eight downstream tasks involving seven evaluation metrics.}
\label{tab:more-abla}
\begin{adjustbox}{width=\textwidth}

\begin{tabular}{lllllllll} 
\toprule
Dataset & Method       & Acc            & M.Pre          & M.Recall       & M.F1           & W.Pre          & W.Recall       & W.F1            \\ 
\midrule
E-K12   & \textbf{FGBERT} & \textbf{0.68}  & \textbf{0.68}  & \textbf{0.61}  & \textbf{0.61}  & \textbf{0.77}  & \textbf{0.67}  & \textbf{0.65}   \\
        & w/o. MGM        & 0.58(↓14.71\%) & 0.56(↓17.65\%) & 0.53(↓13.11\%) & 0.50(↓18.03\%) & 0.58(↓24.68\%) & 0.57(↓14.93\%) & 0.56(↓13.85\%)  \\
        & w/o. Triplet    & 0.61(↓10.29\%) & 0.60(↓11.76\%) & 0.56(↓8.20\%)  & 0.54(↓11.48\%) & 0.61(↓20.78\%) & 0.63(↓5.97\%)  & 0.59(↓9.23\%)   \\ 
\hline
CARD-A  & \textbf{FGBERT} & \textbf{0.91}  & \textbf{0.77}  & \textbf{0.8}   & \textbf{0.78}  & \textbf{0.9}   & \textbf{0.91}  & \textbf{0.9}    \\
        & w/o. MGM        & 0.80(↓12.09\%) & 0.73(↓5.19\%)  & 0.76(↓5.00\%)  & 0.72(↓7.69\%)  & 0.87(↓3.33\%)  & 0.80(↓12.09\%) & 0.79(↓12.22\%)  \\
        & w/o. Triplet    & 0.87(↓4.40\%)  & 0.72(↓6.49\%)  & 0.75(↓6.25\%)  & 0.73(↓6.41\%)  & 0.86(↓4.44\%)  & 0.87(↓4.40\%)  & 0.88(↓2.22\%)   \\ 
\hline
CARD-D  & \textbf{FGBERT} & \textbf{0.85}  & \textbf{0.56}  & \textbf{0.6}   & \textbf{0.57}  & \textbf{0.85}  & \textbf{0.85}  & \textbf{0.85}   \\
        & w/o. MGM        & 0.80(↓5.88\%)  & 0.48(↓14.29\%) & 0.50(↓16.67\%) & 0.47(↓17.54\%) & 0.81(↓4.71\%)  & 0.80(↓5.88\%)  & 0.81(↓4.71\%)   \\
        & w/o. Triplet    & 0.82(↓3.53\%)  & 0.49(↓12.50\%) & 0.58(↓3.33\%)  & 0.52(↓8.77\%)  & 0.82(↓3.53\%)  & 0.82(↓3.53\%)  & 0.82(↓3.53\%)   \\ 
\hline
CARD-R  & \textbf{FGBERT} & \textbf{0.92}  & \textbf{0.75}  & \textbf{0.67}  & \textbf{0.69}  & \textbf{0.91}  & \textbf{0.92}  & \textbf{0.91}   \\
        & w/o. MGM        & 0.88(↓4.35\%)  & 0.65(↓13.33\%) & 0.64(↓4.48\%)  & 0.63(↓8.70\%)  & 0.89(↓2.20\%)  & 0.90(↓2.17\%)  & 0.89(↓2.20\%)   \\
        & w/o. Triplet    & 0.90(↓2.17\%)  & 0.67(↓10.67\%) & 0.65(↓2.99\%)  & 0.66(↓4.35\%)  & 0.89(↓2.20\%)  & 0.9(↓2.17\%)   & 0.89(↓2.20\%)   \\ 
\hline
VFDB    & \textbf{FGBERT} & \textbf{0.91}  & \textbf{0.83}  & \textbf{0.71}  & \textbf{0.75}  & \textbf{0.91}  & \textbf{0.91}  & \textbf{0.91}   \\
        & w/o. MGM        & 0.85(↓6.59\%)  & 0.7(↓15.66\%)  & 0.58(↓18.31\%) & 0.60(↓20.00\%) & 0.85(↓6.59\%)  & 0.84(↓7.69\%)  & 0.85(↓6.59\%)   \\
        & w/o. Triplet    & 0.87(↓4.40\%)  & 0.81(↓2.41\%)  & 0.66(↓7.04\%)  & 0.70(↓6.67\%)  & 0.88(↓3.30\%)  & 0.89(↓2.20\%)  & 0.88(↓3.30\%)   \\ 
\hline
ENZYME  & \textbf{FGBERT} & \textbf{0.99}  & \textbf{0.99}  & \textbf{0.98}  & \textbf{0.99}  & \textbf{0.99}  & \textbf{0.98}  & \textbf{0.99}   \\
        & w/o. MGM        & 0.95(↓4.04\%)  & 0.96(↓3.03\%)  & 0.95(↓3.06\%)  & 0.96(↓3.03\%)  & 0.95(↓4.04\%)  & 0.95(↓3.06\%)  & 0.95(↓4.04\%)   \\
        & w/o. Triplet    & 0.95(↓4.04\%)  & 0.97(↓2.02\%)  & 0.96(↓2.04\%)  & 0.96(↓3.03\%)  & 0.97(↓\%2.02)  & 0.96(↓2.04\%)  & 0.96(↓3.03\%)   \\ 
\hline
PATRIC  & \textbf{FGBERT} & \textbf{0.95}  & \textbf{0.63}  & \textbf{0.68}  & \textbf{0.64}  & \textbf{0.92}  & \textbf{0.94}  & \textbf{0.93}   \\
        & w/o. MGM        & 0.63(↓33.68\%) & 0.34(↓46.03\%) & 0.28(↓58.82\%) & 0.30(↓53.13\%) & 0.67(↓27.17\%) & 0.63(↓32.98\%) & 0.64(↓31.18\%)  \\
        & w/o. Triplet    & 0.83(↓12.63\%) & 0.58(↓7.94\%)  & 0.57(↓16.18\%) & 0.56(↓12.50\%) & 0.81(↓11.96\%) & 0.83(↓11.70\%) & 0.81(↓12.90\%)  \\ 
\hline
NCycDB  & \textbf{FGBERT} & \textbf{0.99}  & \textbf{0.99}  & \textbf{0.99}  & \textbf{0.99}  & \textbf{0.99}  & \textbf{0.99}  & \textbf{0.99}   \\
        & w/o. MGM        & 0.92(↓7.07\%)  & 0.94(↓5.05\%)  & 0.95(↓4.04\%)  & 0.91(↓8.08\%)  & 0.93(↓6.06\%)  & 0.95(↓4.04\%)  & 0.97(↓2.02\%)   \\
        & w/o. Triplet    & 0.95(↓4.04\%)  & 0.96(↓3.03\%)  & 0.96(↓3.03\%)  & 0.95(↓4.04\%)  & 0.97(↓2.02\%)  & 0.96(↓3.03\%)  & 0.97(↓2.02\%)   \\

\bottomrule
\end{tabular}
\end{adjustbox}
\end{table}

\begin{table}[ht]
\centering
\footnotesize
\caption{The ablation study using protein-based gene representation as a context-aware tokenizer.}
\label{tab:app-ablation-tokenizer}
\begin{tabular}{ll|lllllll}
        \toprule
        Dataset & Method & Acc  & M.Pre & M.Recall & M.F1 & W.Pre & W.Recall & W.F1 \\
        \midrule
E-K12   & \textbf{FGBERT} & \textbf{0.68} & \textbf{0.68} & \textbf{0.61} & \textbf{0.61} & \textbf{0.77} & \textbf{0.67} & \textbf{0.65}  \\
        & ESM2            & 0.54          & 0.51          & 0.48          & 0.49          & 0.61          & 0.51          & 0.52           \\
        & BPE             & 0.59          & 0.58          & 0.55          & 0.54          & 0.67          & 0.54          & 0.55           \\ 
\hline
CARD-A  & \textbf{FGBERT} & \textbf{0.91} & \textbf{0.77} & \textbf{0.8}  & \textbf{0.78} & \textbf{0.9}  & \textbf{0.91} & \textbf{0.9}   \\
        & ESM2            & 0.82          & 0.74          & 0.76          & 0.73          & 0.87          & 0.82          & 0.81           \\
        & BPE             & 0.84          & 0.75          & 0.77          & 0.75          & 0.88          & 0.86          & 0.86           \\ 
\hline
CARD-D  & \textbf{FGBERT} & \textbf{0.85} & \textbf{0.56} & \textbf{0.6}  & \textbf{0.57} & \textbf{0.85} & \textbf{0.85} & \textbf{0.85}  \\
        & ESM2            & 0.82          & 0.49          & 0.58          & 0.52          & 0.83          & 0.82          & 0.82           \\
        & BPE             & 0.82          & 0.49          & 0.58          & 0.52          & 0.83          & 0.82          & 0.82           \\ 
\hline
CARD-R  & \textbf{FGBERT} & \textbf{0.92} & \textbf{0.75} & \textbf{0.67} & \textbf{0.69} & \textbf{0.91} & \textbf{0.92} & \textbf{0.91}  \\
        & ESM2            & 0.88          & 0.67          & 0.66          & 0.67          & 0.87          & 0.88          & 0.89           \\
        & BPE             & 0.9           & 0.69          & 0.67          & 0.68          & 0.9           & 0.9           & 0.9            \\ 
\hline
VFDB    & \textbf{FGBERT} & \textbf{0.91} & \textbf{0.83} & \textbf{0.71} & \textbf{0.75} & \textbf{0.91} & \textbf{0.91} & \textbf{0.9}   \\
        & ESM2            & 0.85          & 0.69          & 0.57          & 0.6           & 0.85          & 0.85          & 0.84           \\
        & BPE             & 0.86          & 0.7           & 0.56          & 0.59          & 0.86          & 0.86          & 0.85           \\ 
\hline
ENZYME  & \textbf{FGBERT} & \textbf{0.99} & \textbf{0.99} & \textbf{0.98} & \textbf{0.99} & \textbf{0.99} & \textbf{0.98} & \textbf{0.98}  \\
        & ESM2            & 0.95          & 0.96          & 0.95          & 0.96          & 0.95          & 0.95          & 0.95           \\
        & BPE             & 0.96          & 0.96          & 0.95          & 0.96          & 0.96          & 0.96          & 0.96           \\ 
\hline
PATRIC  & \textbf{FGBERT} & \textbf{0.95} & \textbf{0.63} & \textbf{0.68} & \textbf{0.99} & \textbf{0.92} & \textbf{0.94} & \textbf{0.99}  \\
        & ESM2            & 0.63          & 0.34          & 0.27          & 0.3           & 0.63          & 0.61          & 0.62           \\
        & BPE             & 0.56          & 0.28          & 0.28          & 0.27          & 0.6           & 0.56          & 0.57           \\ 
\hline
NCycDB  & \textbf{FGBERT} & \textbf{0.99} & \textbf{0.99} & \textbf{0.99} & \textbf{0.99} & \textbf{0.99} & \textbf{0.99} & \textbf{0.99}  \\
        & ESM2            & 0.56          & 0.28          & 0.28          & 0.27          & 0.6           & 0.56          & 0.57           \\
        & BPE             & 0.63          & 0.33          & 0.27          & 0.29          & 0.66          & 0.62          & 0.63           \\

\bottomrule
\end{tabular}
\end{table}
We have replaced the original tokenizer with ESM2 representations and BPE and assessed its impact on various downstream tasks.
Tab.~\ref{tab:app-ablation-tokenizer} demonstrates that the FGBERT tokenizer outperforms ESM2 and BPE representations across various metrics, indicating its efficacy in capturing intricate relationships between gene sequences and their functions. While DNABERT-2's use of BPE efficiently compresses sequences, our context-aware tokenizer excels in precision with metagenomic data.

\subsection{Detailed Classification Results on CARD Dataset}
This section provides more detailed ablation and analytical experiments on the individual modules.

\begin{table}[ht]
\centering
\caption{Detailed Classification Results of {FGBERT} Model on Drug Class Category Prediction.}
\label{tab:CARD_ours_Drug_ratio}
\begin{tabular}{p{3.5cm}p{0.8cm}p{0.8cm}p{0.8cm}}
\toprule
Class Name                          & Total Count & Correctly Num & Correct Ratio \\ \midrule
cephalosporin                       & 103         & 101           & 98.06\%       \\
aminoglycoside antibiotic           & 47          & 37            & 78.72\%       \\
cephamycin                          & 34          & 31            & 91.18\%       \\
fluoroquinolone antibiotic          & 33          & 27            & 81.82\%       \\
carbapenem                          & 31          & 25            & 80.65\%       \\
peptide antibiotic                  & 30          & 26            & 86.67\%       \\
penam                               & 23          & 21            & 91.30\%       \\
tetracycline antibiotic             & 16          & 10            & 62.50\%       \\
macrolide antibiotic                & 12          & 10            & 83.33\%       \\
diaminopyrimidine antibiotic        & 11          & 8             & 72.73\%       \\
phenicol antibiotic                 & 11          & 6             & 54.55\%       \\
phosphonic acid antibiotic          & 10          & 6             & 60.00\%       \\
lincosamide antibiotic              & 3           & 0             & 0.00          \\
rifamycin antibiotic                & 3           & 1             & 33.33\%       \\
disinfecting agents antiseptics     & 3           & 2             & 66.67\%       \\
nitroimidazole antibiotic           & 3           & 1             & 33.33\%       \\
aminocoumarin antibiotic            & 2           & 0             & 0.00\%          \\
sulfonamide antibiotic              & 2           & 0             & 0.00\%         \\
elfamycin antibiotic                & 2           & 2             & 100.00\%         \\
isoniazid-like antibiotic           & 2           & 1             & 50.00\%       \\
fusidane antibiotic                 & 1           & 0             & 0.00\%          \\
nucleoside antibiotic               & 1           & 0             & 0.00\%          \\
polyamine antibiotic                & 1           & 0             & 0.00\%          \\
thioamide antibiotic                & 1           & 0             & 0.00\%          \\
glycopeptide antibiotic             & 1           & 0             & 0.00\%          \\
mupirocin-like antibiotic           & 1           & 0             & 0.00\%          \\
pleuromutilin antibiotic            & 1           & 0             & 0.00\%          \\
salicylic acid antibiotic           & 1           & 0             & 0.00\%          \\ \bottomrule
\end{tabular}
\end{table}

\begin{footnotesize}   
\begin{longtable}{p{10cm}p{0.5cm}p{0.5cm}p{1cm}}

\caption{Detailed Classification Results of {FGBERT} on AMR Gene Family Category Prediction.}  
\label{tab:CARD_ours_AMR_ratio} \\
\toprule
Class Name & Total Count & Correct Num & Correct Ratio \\ 
\midrule
\endfirsthead
\multicolumn{4}{c}%
{{\bfseries \tablename\ \thetable{} -- continued from previous page}} \\
\hline
Class Name & Total Count & Correct Num & Correct Ratio \\ 
\midrule
\endhead
\hline \multicolumn{4}{r}{{Continued on next page}} \\ 
\endfoot
\bottomrule
\endlastfoot

CTX-M beta-lactamase                                                             & 46          & 46          & 100.00\%      \\
ADC beta-lactamases pending classification for carbapenemase activity            & 36          & 36          & 100.00\%      \\
CMY beta-lactamase                                                               & 33          & 33          & 100.00\%      \\
quinolone resistance protein (qnr)                                               & 22          & 22          & 100.00\%      \\
MCR phosphoethanolamine transferase                                              & 21          & 20          & 95.24\%       \\
major facilitator superfamily (MFS) antibiotic efflux pump                       & 17          & 15          & 88.24\%       \\
trimethoprim resistant dihydrofolate reductase dfr                               & 12          & 8           & 66.67\%       \\
AAC(6')                                                                          & 12          & 12          & 100.00\%      \\
CARB beta-lactamase                                                              & 10          & 9           & 90.00\%       \\
ADC beta-lactamase without carbapenemase activity                                & 10          & 3           & 30.00\%       \\
16s rRNA with mutation conferring resistance to aminoglycoside antibiotics       & 9           & 7           & 77.78\%       \\
fosfomycin thiol transferase                                                     & 6           & 3           & 50.00\%       \\
ANT(3'')                                                                         & 6           & 6           & 100.00\%      \\
chloramphenicol acetyltransferase (CAT)                                          & 6           & 5           & 83.33\%       \\
23S rRNA with mutation conferring resistance to macrolide antibiotics            & 5           & 5           & 100.00\%      \\
APH(3')                                                                          & 4           & 2           & 50.00\%       \\
AAC(3)                                                                           & 4           & 4           & 100.00\%      \\
IND beta-lactamase                                                               & 4           & 4           & 100.00\%      \\
ATP-binding cassette (ABC) antibiotic efflux pump                                & 4           & 3           & 75.00\%       \\
tetracycline-resistant ribosomal protection protein                              & 4           & 3           & 75.00\%       \\
IMI beta-lactamase                                                               & 4           & 4           & 100.00\%      \\
CfiA beta-lactamase                                                              & 4           & 4           & 100.00\%      \\
fluoroquinolone resistant gyrA                                                   & 4           & 2           & 50.00\%       \\
16S rRNA methyltransferase (G1405)                                               & 3           & 1           & 33.33\%       \\
resistance-nodulation-cell division(RND) antibiotic efflux pump                  & 3           & 0           & 0.00\%        \\
macrolide phosphotransferase (MPH)                                               & 3           & 0           & 0.00\%        \\
tetracycline inactivation enzyme                                                 & 3           & 3           & 100.00\%      \\
fluoroquinolone resistant parC                                                   & 3           & 3           & 100.00\%      \\
lincosamide nucleotidyltransferase (LNU)                                         & 2           & 1           & 50.00\%       \\
rifampin ADP-ribosyltransferase (Arr)                                            & 2           & 2           & 100.00\%      \\
CphA beta-lactamase                                                              & 2           & 2           & 100.00\%      \\
AAC(2')                                                                          & 2           & 2           & 100.00\%      \\
FRI beta-lactamase                                                               & 2           & 2           & 100.00\%      \\
HERA beta-lactamase                                                              & 2           & 2           & 100.00\%      \\
EC beta-lactamase                                                                & 2           & 2           & 100.00\%      \\
nitroimidazole reductase                                                         & 2           & 1           & 50.00\%       \\
Target protecting FusB-type protein conferring resistance to Fusidic acid        & 1           & 1           & 100.00\%      \\
SME beta-lactamase                                                               & 1           & 1           & 100.00\%      \\
Miscellaneous ABC-F subfamily ATP-binding cassette ribosomal protection proteins & 1           & 0           & 0.00\%        \\
APH(2'')                                                                         & 1           & 0           & 0.00\%        \\
CfxA beta-lactamase                                                              & 1           & 1           & 100.00\%      \\
SRT beta-lactamase                                                               & 1           & 1           & 100.00\%      \\
aminoglycoside bifunctional resistance protein                                   & 1           & 0           & 0.00\%        \\
APH(9)                                                                           & 1           & 0           & 0.00\%        \\
sulfonamide resistant sul                                                        & 1           & 0           & 0.00\%        \\
ANT(6)                                                                           & 1           & 0           & 0.00\%        \\
multidrug and toxic compound extrusion (MATE) transporter                        & 1           & 0           & 0.00\%        \\
methicillin resistant PBP2                                                       & 1           & 1           & 100.00\%      \\
streptothricin acetyltransferase (SAT)                                           & 1           & 0           & 0.00\%        \\
pmr phosphoethanolamine transferase                                              & 1           & 1           & 100.00\%      \\
BlaZ beta-lactamase                                                              & 1           & 0           & 0.00\%        \\
rifampin monooxygenase                                                           & 1           & 0           & 0.00\%        \\
APH(6)                                                                           & 1           & 0           & 0.00\%        \\
macrolide esterase                                                               & 1           & 0           & 0.00\%        \\
ANT(4')                                                                          & 1           & 0           & 0.00\%        \\
ANT(9)                                                                           & 1           & 0           & 0.00\%        \\
AQU beta-lactamase                                                               & 1           & 1           & 100.00\%      \\
CepA beta-lactamase                                                              & 1           & 1           & 100.00\%      \\
APH(3'')                                                                         & 1           & 0           & 0.00\%        \\
sulfonamide resistant dihydropteroate synthase folP                              & 1           & 0           & 0.00\%        \\
aminocoumarin resistant gyrB                                                     & 1           & 0           & 0.00\%        \\
fluoroquinolone resistant parE                                                   & 1           & 0           & 0.00\%        \\
daptomycin resistant cls                                                         & 1           & 0           & 0.00\%        \\
16s rRNA with mutation conferring resistance to peptide antibiotics              & 1           & 0           & 0.00\%        \\
elfamycin resistant EF-Tu                                                        & 1           & 0           & 0.00\%        \\
defensin resistant mprF                                                          & 1           & 1           & 100.00\%      \\
antibiotic resistant ndh                                                         & 1           & 0           & 0.00\%        \\
fluoroquinolone resistant gyrB                                                   & 1           & 0           & 0.00\%        \\
16S rRNA with mutation conferring resistance to tetracycline derivatives         & 1           & 0           & 0.00\%        \\
FONA beta-lactamase                                                              & 1           & 1           & 100.00\%      \\
antibiotic-resistant isoleucyl-tRNA   synthetase (ileS)                          & 1           & 1           & 100.00\%      \\
antibiotic-resistant murA transferase                                            & 1           & 0           & 0.00\%        \\
antibiotic-resistant UhpT                                                        & 1           & 1           & 100.00\%      \\
small multidrug resistance (SMR)antibiotic efflux pump                           & 1           & 0           & 0.00\%        \\
ARL Beta-lactamase                                                               & 1           & 1           & 100.00\%      \\
General Bacterial Porin with reduced permeability to peptide antibiotics         & 1           & 0           & 0.00\%        \\
CMH beta-lactamase                                                               & 1           & 1           & 100.00\%      \\
AXC beta-lactamase                                                               & 1           & 1           & 100.00\%      \\
CRH beta-lactamase                                                               & 1           & 1           & 100.00\%      \\
KLUC beta-lactamase                                                              & 1           & 1           & 100.00\%      \\
LUT beta-lactamase                                                               & 1           & 1           & 100.00\%      \\
PFM beta-lactamase                                                               & 1           & 1           & 100.00\%      \\
SGM beta-lactamase                                                               & 1           & 1           & 100.00\%      \\
\end{longtable}
\end{footnotesize}

\section{Broader Impacts}
In our work, we focus on the wider application of macro-genomic pre-training models, particularly in the field of functional annotation. We argue that the positive societal impacts of such models include, but are not limited to, improved disease surveillance and prediction, facilitation of new drug development and optimisation of vaccine design. However, we should also be aware of possible negative impacts such as potential risk of misuse and privacy concerns.

\section{Prelinaries.}
\input{sec_preliminaries}

%% file: sec_preliminaries.tex
Transformers are the dominant tools for modeling sequence data.
In Transformer, the input sequence is first transformed into a series of vectors, each corresponding to an element of the sequence (e.g., a word or character), which are generated by an embedding layer.

Subsequently, self-attention mechanism computes the attention score of each element with respect to the other elements in the sequence.
Specifically, by multiplying three different weight matrices, each input vector $X$ is transformed into: a query vector $Q$, a key vector $K$, and a value vector $V$. Then, the attention score is computed.
\begin{equation}
    \textit{Attention}(Q,K)=\frac{Q\cdot K^T}{\sqrt{d_k}}
\end{equation}
These scores are normalized by Softmax function, where it is multiplied by the value vector $V$ to get the weighted value vector. All these weighted value vectors are summed to form the output Z of the attention layer.
\begin{equation}
    Z = \textit{softmax} (\textit{Attention}(Q,K))V
\end{equation}
Transformers typically use a multi-head attention mechanism to capture the information in the sequence more comprehensively.
The outputs of each self-attention and feed-forward network layer are passed through a residual connection, followed by layer normalization.
After multi-layer processing, the Transformer model outputs a series of vectors suitable for a variety of downstream tasks such as categorization, translation, or text generation of text, speech, or any form of sequence data.

%% file: main_nips24_fgbert.bbl
\begin{thebibliography}{10}

\bibitem{uniprot2023uniprot}
Uniprot: the universal protein knowledgebase in 2023.
\newblock {\em Nucleic Acids Research}, 51(D1):D523--D531, 2023.

\bibitem{al2019cnn}
Amani Al-Ajlan and Achraf El~Allali.
\newblock Cnn-mgp: convolutional neural networks for metagenomics gene prediction.
\newblock {\em Interdisciplinary Sciences: Computational Life Sciences}, 11:628--635, 2019.

\bibitem{al2022diverse}
Basem Al-Shayeb, Petr Skopintsev, Katarzyna~M Soczek, Elizabeth~C Stahl, Zheng Li, Evan Groover, Dylan Smock, Amy~R Eggers, Patrick Pausch, Brady~F Cress, et~al.
\newblock Diverse virus-encoded crispr-cas systems include streamlined genome editors.
\newblock {\em Cell}, 185(24):4574--4586, 2022.

\bibitem{albertsen2023long}
Mads Albertsen.
\newblock Long-read metagenomics paves the way toward a complete microbial tree of life.
\newblock {\em Nature Methods}, 20(1):30--31, 2023.

\bibitem{arnold2022horizontal}
Brian~J Arnold, I-Ting Huang, and William~P Hanage.
\newblock Horizontal gene transfer and adaptive evolution in bacteria.
\newblock {\em Nature Reviews Microbiology}, 20(4):206--218, 2022.

\bibitem{avsec2021effective}
{\v{Z}}iga Avsec, Vikram Agarwal, Daniel Visentin, Joseph~R Ledsam, Agnieszka Grabska-Barwinska, Kyle~R Taylor, Yannis Assael, John Jumper, Pushmeet Kohli, and David~R Kelley.
\newblock Effective gene expression prediction from sequence by integrating long-range interactions.
\newblock {\em Nature methods}, 18(10):1196--1203, 2021.

\bibitem{bairoch2000enzyme}
Amos Bairoch.
\newblock The enzyme database in 2000.
\newblock {\em Nucleic acids research}, 28(1):304--305, 2000.

\bibitem{chen2005vfdb}
Lihong Chen, Jian Yang, Jun Yu, Zhijian Yao, Lilian Sun, Yan Shen, and Qi~Jin.
\newblock Vfdb: a reference database for bacterial virulence factors.
\newblock {\em Nucleic acids research}, 33(suppl\_1):D325--D328, 2005.

\bibitem{cui2007micrornas}
Qinghua Cui, Zhenbao Yu, Youlian Pan, Enrico~O Purisima, and Edwin Wang.
\newblock Micrornas preferentially target the genes with high transcriptional regulation complexity.
\newblock {\em Biochemical and biophysical research communications}, 352(3):733--738, 2007.

\bibitem{dalla2023nucleotide}
Hugo Dalla-Torre, Liam Gonzalez, Javier Mendoza-Revilla, Nicolas~Lopez Carranza, Adam~Henryk Grzywaczewski, Francesco Oteri, Christian Dallago, Evan Trop, Bernardo~P de~Almeida, Hassan Sirelkhatim, et~al.
\newblock The nucleotide transformer: Building and evaluating robust foundation models for human genomics.
\newblock {\em bioRxiv}, pages 2023--01, 2023.

\bibitem{devlin2018bert}
Jacob Devlin, Ming-Wei Chang, Kenton Lee, and Kristina Toutanova.
\newblock Bert: Pre-training of deep bidirectional transformers for language understanding.
\newblock {\em arXiv preprint arXiv:1810.04805}, 2018.

\bibitem{d2014redundancy}
David~J D'Onofrio and David~L Abel.
\newblock Redundancy of the genetic code enables translational pausing.
\newblock {\em Frontiers in genetics}, 5:140, 2014.

\bibitem{ermolaeva2001prediction}
Maria~D Ermolaeva, Owen White, and Steven~L Salzberg.
\newblock Prediction of operons in microbial genomes.
\newblock {\em Nucleic acids research}, 29(5):1216--1221, 2001.

\bibitem{fiannaca2018deep}
Antonino Fiannaca, Laura La~Paglia, Massimo La~Rosa, Giosue’ Lo~Bosco, Giovanni Renda, Riccardo Rizzo, Salvatore Gaglio, and Alfonso Urso.
\newblock Deep learning models for bacteria taxonomic classification of metagenomic data.
\newblock {\em BMC bioinformatics}, 19:61--76, 2018.

\bibitem{gillespie2011patric}
Joseph~J Gillespie, Alice~R Wattam, Stephen~A Cammer, Joseph~L Gabbard, Maulik~P Shukla, Oral Dalay, Timothy Driscoll, Deborah Hix, Shrinivasrao~P Mane, Chunhong Mao, et~al.
\newblock Patric: the comprehensive bacterial bioinformatics resource with a focus on human pathogenic species.
\newblock {\em Infection and immunity}, 79(11):4286--4298, 2011.

\bibitem{gruber2008earth}
Nicolas Gruber and James~N Galloway.
\newblock An earth-system perspective of the global nitrogen cycle.
\newblock {\em Nature}, 451(7176):293--296, 2008.

\bibitem{gruenstaeudl2020annonex2embl}
Michael Gruenstaeudl.
\newblock annonex2embl: automatic preparation of annotated dna sequences for bulk submissions to ena.
\newblock {\em Bioinformatics}, 36(12):3841--3848, 2020.

\bibitem{gwak2022vibe}
Ho-Jin Gwak and Mina Rho.
\newblock Vibe: a hierarchical bert model to identify eukaryotic viruses using metagenome sequencing data.
\newblock {\em Briefings in Bioinformatics}, 23(4):bbac204, 2022.

\bibitem{hermans2017defense}
Alexander Hermans, Lucas Beyer, and Bastian Leibe.
\newblock In defense of the triplet loss for person re-identification.
\newblock {\em arXiv preprint arXiv:1703.07737}, 2017.

\bibitem{hoarfrost2022deep}
A~Hoarfrost, A~Aptekmann, G~Farfa{\~n}uk, and Y~Bromberg.
\newblock Deep learning of a bacterial and archaeal universal language of life enables transfer learning and illuminates microbial dark matter.
\newblock {\em Nature communications}, 13(1):2606, 2022.

\bibitem{hochreiter1997long}
Sepp Hochreiter and J{\"u}rgen Schmidhuber.
\newblock Long short-term memory.
\newblock {\em Neural computation}, 9(8):1735--1780, 1997.

\bibitem{hong2008atp}
Sangjin Hong and Peter~L Pedersen.
\newblock Atp synthase and the actions of inhibitors utilized to study its roles in human health, disease, and other scientific areas.
\newblock {\em Microbiology and molecular biology reviews}, 72(4):590--641, 2008.

\bibitem{hu2022metagenomic}
Yanping Hu, Yangcan Chen, Jing Xu, Xinge Wang, Shengqiu Luo, Bangwei Mao, Qi~Zhou, and Wei Li.
\newblock Metagenomic discovery of novel crispr-cas13 systems.
\newblock {\em Cell Discovery}, 8(1):107, 2022.

\bibitem{jain2017duplication}
Siddharth Jain, Farzad~Farnoud Hassanzadeh, Moshe Schwartz, and Jehoshua Bruck.
\newblock Duplication-correcting codes for data storage in the dna of living organisms.
\newblock {\em IEEE Transactions on Information Theory}, 63(8):4996--5010, 2017.

\bibitem{ji2021dnabert}
Yanrong Ji, Zhihan Zhou, Han Liu, and Ramana~V Davuluri.
\newblock Dnabert: pre-trained bidirectional encoder representations from transformers model for dna-language in genome.
\newblock {\em Bioinformatics}, 37(15):2112--2120, 2021.

\bibitem{jia2016card}
Baofeng Jia, Amogelang~R Raphenya, Brian Alcock, Nicholas Waglechner, Peiyao Guo, Kara~K Tsang, Briony~A Lago, Biren~M Dave, Sheldon Pereira, Arjun~N Sharma, et~al.
\newblock Card 2017: expansion and model-centric curation of the comprehensive antibiotic resistance database.
\newblock {\em Nucleic acids research}, page gkw1004, 2016.

\bibitem{jian2021antibiotic}
Zonghui Jian, Li~Zeng, Taojie Xu, Shuai Sun, Shixiong Yan, Lan Yang, Ying Huang, Junjing Jia, and Tengfei Dou.
\newblock Antibiotic resistance genes in bacteria: Occurrence, spread, and control.
\newblock {\em Journal of basic microbiology}, 61(12):1049--1070, 2021.

\bibitem{jones2008global}
Kate~E Jones, Nikkita~G Patel, Marc~A Levy, Adam Storeygard, Deborah Balk, John~L Gittleman, and Peter Daszak.
\newblock Global trends in emerging infectious diseases.
\newblock {\em Nature}, 451(7181):990--993, 2008.

\bibitem{jumper2021highly}
John Jumper, Richard Evans, Alexander Pritzel, Tim Green, Michael Figurnov, Olaf Ronneberger, Kathryn Tunyasuvunakool, Russ Bates, Augustin {\v{Z}}{\'\i}dek, Anna Potapenko, et~al.
\newblock Highly accurate protein structure prediction with alphafold.
\newblock {\em Nature}, 596(7873):583--589, 2021.

\bibitem{karp2019biocyc}
Peter~D Karp, Richard Billington, Ron Caspi, Carol~A Fulcher, Mario Latendresse, Anamika Kothari, Ingrid~M Keseler, Markus Krummenacker, Peter~E Midford, Quang Ong, et~al.
\newblock The biocyc collection of microbial genomes and metabolic pathways.
\newblock {\em Briefings in bioinformatics}, 20(4):1085--1093, 2019.

\bibitem{khosla2020supervised}
Prannay Khosla, Piotr Teterwak, Chen Wang, Aaron Sarna, Yonglong Tian, Phillip Isola, Aaron Maschinot, Ce~Liu, and Dilip Krishnan.
\newblock Supervised contrastive learning.
\newblock {\em Advances in neural information processing systems}, 33:18661--18673, 2020.

\bibitem{KUSTERS19934119}
Karl~A. Kusters, Sotiris~E. Pratsinis, Steven~G. Thoma, and Douglas~M. Smith.
\newblock Ultrasonic fragmentation of agglomerate powders.
\newblock {\em Chemical Engineering Science}, 48(24):4119--4127, 1993.

\bibitem{lawson2004catabolite}
Catherine~L Lawson, David Swigon, Katsuhiko~S Murakami, Seth~A Darst, Helen~M Berman, and Richard~H Ebright.
\newblock Catabolite activator protein: Dna binding and transcription activation.
\newblock {\em Current opinion in structural biology}, 14(1):10--20, 2004.

\bibitem{lee2022multimodal}
Seung~Jae Lee and Mina Rho.
\newblock Multimodal deep learning applied to classify healthy and disease states of human microbiome.
\newblock {\em Scientific Reports}, 12(1):824, 2022.

\bibitem{liang2020deepmicrobes}
Qiaoxing Liang, Paul~W Bible, Yu~Liu, Bin Zou, and Lai Wei.
\newblock Deepmicrobes: taxonomic classification for metagenomics with deep learning.
\newblock {\em NAR Genomics and Bioinformatics}, 2(1):lqaa009, 2020.

\bibitem{lin2022language}
Zeming Lin, Halil Akin, Roshan Rao, Brian Hie, Zhongkai Zhu, Wenting Lu, Nikita Smetanin, Allan dos Santos~Costa, Maryam Fazel-Zarandi, Tom Sercu, Sal Candido, et~al.
\newblock Language models of protein sequences at the scale of evolution enable accurate structure prediction.
\newblock {\em bioRxiv}, 2022.

\bibitem{lin2023evolutionary}
Zeming Lin, Halil Akin, Roshan Rao, Brian Hie, Zhongkai Zhu, Wenting Lu, Nikita Smetanin, Robert Verkuil, Ori Kabeli, Yaniv Shmueli, et~al.
\newblock Evolutionary-scale prediction of atomic-level protein structure with a language model.
\newblock {\em Science}, 379(6637):1123--1130, 2023.

\bibitem{liu2022opportunities}
Sijia Liu, Christina~D Moon, Nan Zheng, Sharon Huws, Shengguo Zhao, and Jiaqi Wang.
\newblock Opportunities and challenges of using metagenomic data to bring uncultured microbes into cultivation.
\newblock {\em Microbiome}, 10(1):76, 2022.

\bibitem{loshchilov2017decoupled}
Ilya Loshchilov and Frank Hutter.
\newblock Decoupled weight decay regularization.
\newblock {\em arXiv preprint arXiv:1711.05101}, 2017.

\bibitem{lu2022machine}
Hongyuan Lu, Daniel~J Diaz, Natalie~J Czarnecki, Congzhi Zhu, Wantae Kim, Raghav Shroff, Daniel~J Acosta, Bradley~R Alexander, Hannah~O Cole, Yan Zhang, et~al.
\newblock Machine learning-aided engineering of hydrolases for pet depolymerization.
\newblock {\em Nature}, 604(7907):662--667, 2022.

\bibitem{mande2012classification}
Sharmila~S Mande, Monzoorul~Haque Mohammed, and Tarini~Shankar Ghosh.
\newblock Classification of metagenomic sequences: methods and challenges.
\newblock {\em Briefings in bioinformatics}, 13(6):669--681, 2012.

\bibitem{martinez2002interactions}
Jos{\'e}~L Mart{\'\i}nez and Fernando Baquero.
\newblock Interactions among strategies associated with bacterial infection: pathogenicity, epidemicity, and antibiotic resistance.
\newblock {\em Clinical microbiology reviews}, 15(4):647--679, 2002.

\bibitem{mathieu2022machine}
Alban Mathieu, Mickael Leclercq, Melissa Sanabria, Olivier Perin, and Arnaud Droit.
\newblock Machine learning and deep learning applications in metagenomic taxonomy and functional annotation.
\newblock {\em Frontiers in Microbiology}, 13:811495, 2022.

\bibitem{mcarthur2013comprehensive}
Andrew~G McArthur, Nicholas Waglechner, Fazmin Nizam, Austin Yan, Marisa~A Azad, Alison~J Baylay, Kirandeep Bhullar, Marc~J Canova, Gianfranco De~Pascale, Linda Ejim, et~al.
\newblock The comprehensive antibiotic resistance database.
\newblock {\em Antimicrobial agents and chemotherapy}, 57(7):3348--3357, 2013.

\bibitem{mcwilliam2013analysis}
Hamish McWilliam, Weizhong Li, Mahmut Uludag, Silvano Squizzato, Young~Mi Park, Nicola Buso, Andrew~Peter Cowley, and Rodrigo Lopez.
\newblock Analysis tool web services from the embl-ebi.
\newblock {\em Nucleic acids research}, 41(W1):W597--W600, 2013.

\bibitem{miao2022virtifier}
Yan Miao, Fu~Liu, Tao Hou, and Yun Liu.
\newblock Virtifier: a deep learning-based identifier for viral sequences from metagenomes.
\newblock {\em Bioinformatics}, 38(5):1216--1222, 2022.

\bibitem{miller2022deciphering}
Danielle Miller, Adi Stern, and David Burstein.
\newblock Deciphering microbial gene function using natural language processing.
\newblock {\em Nature Communications}, 13(1):5731, 2022.

\bibitem{nguyen2023hyenadna}
Eric Nguyen, Michael Poli, Marjan Faizi, Armin Thomas, Callum Birch-Sykes, Michael Wornow, Aman Patel, Clayton Rabideau, Stefano Massaroli, Yoshua Bengio, et~al.
\newblock Hyenadna: Long-range genomic sequence modeling at single nucleotide resolution.
\newblock {\em arXiv preprint arXiv:2306.15794}, 2023.

\bibitem{nomenclature1992recommendations}
Enzyme Nomenclature.
\newblock Recommendations of the nomenclature committee of the international union of biochemistry and molecular biology on the nomenclature and classification of enzymes, 1992.

\bibitem{o2016tackling}
Jim O'Neill.
\newblock Tackling drug-resistant infections globally: final report and recommendations.
\newblock 2016.

\bibitem{park2020enhancing}
Sungjoon Park, Yookyung Koh, Hwisang Jeon, Hyunjae Kim, Yoonsun Yeo, and Jaewoo Kang.
\newblock Enhancing the interpretability of transcription factor binding site prediction using attention mechanism.
\newblock {\em Scientific reports}, 10(1):13413, 2020.

\bibitem{pastinen2006influence}
Tomi Pastinen, Bing Ge, and Thomas~J Hudson.
\newblock Influence of human genome polymorphism on gene expression.
\newblock {\em Human molecular genetics}, 15(suppl\_1):R9--R16, 2006.

\bibitem{pavlopoulos2023unraveling}
Georgios~A Pavlopoulos, Fotis~A Baltoumas, Sirui Liu, Oguz Selvitopi, Antonio~Pedro Camargo, Stephen Nayfach, Ariful Azad, Simon Roux, Lee Call, Natalia~N Ivanova, et~al.
\newblock Unraveling the functional dark matter through global metagenomics.
\newblock {\em Nature}, 622(7983):594--602, 2023.

\bibitem{ren2020identifying}
Jie Ren, Kai Song, Chao Deng, Nathan~A Ahlgren, Jed~A Fuhrman, Yi~Li, Xiaohui Xie, Ryan Poplin, and Fengzhu Sun.
\newblock Identifying viruses from metagenomic data using deep learning.
\newblock {\em Quantitative Biology}, 8:64--77, 2020.

\bibitem{salgado2018using}
Heladia Salgado, Irma Mart{\'\i}nez-Flores, V{\'\i}ctor~H Bustamante, Kevin Alquicira-Hern{\'a}ndez, Jair~S Garc{\'\i}a-Sotelo, Delfino Garc{\'\i}a-Alonso, and Julio Collado-Vides.
\newblock Using regulondb, the escherichia coli k-12 gene regulatory transcriptional network database.
\newblock {\em Current protocols in bioinformatics}, 61(1):1--32, 2018.

\bibitem{schuster1997bidirectional}
Mike Schuster and Kuldip~K Paliwal.
\newblock Bidirectional recurrent neural networks.
\newblock {\em IEEE transactions on Signal Processing}, 45(11):2673--2681, 1997.

\bibitem{segata2012metagenomic}
Nicola Segata, Levi Waldron, Annalisa Ballarini, Vagheesh Narasimhan, Olivier Jousson, and Curtis Huttenhower.
\newblock Metagenomic microbial community profiling using unique clade-specific marker genes.
\newblock {\em Nature methods}, 9(8):811--814, 2012.

\bibitem{tawfik2010enzyme}
Olga~Khersonsky Tawfik and Dan S.
\newblock Enzyme promiscuity: a mechanistic and evolutionary perspective.
\newblock {\em Annual review of biochemistry}, 79:471--505, 2010.

\bibitem{tu2019ncycdb}
Qichao Tu, Lu~Lin, Lei Cheng, Ye~Deng, and Zhili He.
\newblock Ncycdb: a curated integrative database for fast and accurate metagenomic profiling of nitrogen cycling genes.
\newblock {\em Bioinformatics}, 35(6):1040--1048, 2019.

\bibitem{wichmann2023metatransformer}
Alexander Wichmann, Etienne Buschong, Andr{\'e} M{\"u}ller, Daniel J{\"u}nger, Andreas Hildebrandt, Thomas Hankeln, and Bertil Schmidt.
\newblock Metatransformer: deep metagenomic sequencing read classification using self-attention models.
\newblock {\em NAR Genomics and Bioinformatics}, 5(3):lqad082, 2023.

\bibitem{wolf2020transformers}
Thomas Wolf, Lysandre Debut, Victor Sanh, Julien Chaumond, Clement Delangue, Anthony Moi, Pierric Cistac, Tim Rault, R{\'e}mi Louf, Morgan Funtowicz, et~al.
\newblock Transformers: State-of-the-art natural language processing.
\newblock In {\em Proceedings of the 2020 conference on empirical methods in natural language processing: system demonstrations}, pages 38--45, 2020.

\bibitem{yan2020deepte}
Haidong Yan, Aureliano Bombarely, and Song Li.
\newblock Deepte: a computational method for de novo classification of transposons with convolutional neural network.
\newblock {\em Bioinformatics}, 36(15):4269--4275, 2020.

\bibitem{yang2021review}
Chao Yang, Debajyoti Chowdhury, Zhenmiao Zhang, William~K Cheung, Aiping Lu, Zhaoxiang Bian, and Lu~Zhang.
\newblock A review of computational tools for generating metagenome-assembled genomes from metagenomic sequencing data.
\newblock {\em Computational and Structural Biotechnology Journal}, 19:6301--6314, 2021.

\bibitem{yang2022integrating}
Meng Yang, Lichao Huang, Haiping Huang, Hui Tang, Nan Zhang, Huanming Yang, Jihong Wu, and Feng Mu.
\newblock Integrating convolution and self-attention improves language model of human genome for interpreting non-coding regions at base-resolution.
\newblock {\em Nucleic acids research}, 50(14):e81--e81, 2022.

\bibitem{yang2016args}
Ying Yang, Xiaotao Jiang, Benli Chai, Liping Ma, Bing Li, Anni Zhang, James~R Cole, James~M Tiedje, and Tong Zhang.
\newblock Args-oap: online analysis pipeline for antibiotic resistance genes detection from metagenomic data using an integrated structured arg-database.
\newblock {\em Bioinformatics}, 32(15):2346--2351, 2016.

\bibitem{zhang2021metagenomics}
Zeng Zhang, Zhe Han, Yuqing Wu, Shuaiming Jiang, Chenchen Ma, Yanjun Zhang, and Jiachao Zhang.
\newblock Metagenomics assembled genome scale analysis revealed the microbial diversity and genetic polymorphism of lactiplantibacillus plantarum in traditional fermented foods of hainan, china.
\newblock {\em Food Research International}, 150:110785, 2021.

\bibitem{zhou2023dnabert}
Zhihan Zhou, Yanrong Ji, Weijian Li, Pratik Dutta, Ramana Davuluri, and Han Liu.
\newblock Dnabert-2: Efficient foundation model and benchmark for multi-species genome.
\newblock {\em arXiv preprint arXiv:2306.15006}, 2023.

\bibitem{zvyagin2022genslms}
Maxim Zvyagin, Alexander Brace, Kyle Hippe, Yuntian Deng, Bin Zhang, Cindy~Orozco Bohorquez, Austin Clyde, Bharat Kale, Danilo Perez-Rivera, Heng Ma, et~al.
\newblock Genslms: Genome-scale language models reveal sars-cov-2 evolutionary dynamics.
\newblock {\em The International Journal of High Performance Computing Applications}, page 10943420231201154, 2022.

\end{thebibliography}
